\documentclass[conference,compsoc]{IEEEtran}
\usepackage[affil-it]{authblk}
\usepackage{amsmath}
\usepackage{amssymb}
\usepackage{array}
\usepackage{balance}
\usepackage{booktabs}
\usepackage[font={small}]{caption}
\usepackage{enumitem}
\usepackage[multiple]{footmisc}
\usepackage{graphicx}
\usepackage[misc]{ifsym}
\usepackage{makecell}
\usepackage{multirow}
\usepackage{subfig}
\usepackage{times}
\usepackage{titlesec}
\usepackage{xcolor}

\makeatletter
\def\@citex[#1]#2{\leavevmode
\let\@citea\@empty
\@cite{\@for\@citeb:=#2\do
{\@citea\def\@citea{,\penalty\@m\ }%
\edef\@citeb{\expandafter\@firstofone\@citeb\@empty}%
\if@filesw\immediate\write\@auxout{\string\citation{\@citeb}}\fi
\@ifundefined{b@\@citeb}{\hbox{\reset@font\bfseries ?}%
\G@refundefinedtrue
\@latex@warning
{Citation `\@citeb' on page \thepage \space undefined}}%
{\@cite@ofmt{\csname b@\@citeb\endcsname}}}}{#1}}
\makeatother

\def\tablename{Table}

\begin{document}

\title{RingAuth: Wearable Authentication using a Smart Ring}
\author{Jack Sturgess, Simon Birnbach, Simon Eberz, and Ivan Martinovic}
\affil{Department of Computer Science, University of Oxford, Oxford, UK \\\{firstname.lastname\}@cs.ox.ac.uk}
\maketitle

\begin{abstract}
In this paper, we show that by using inertial sensor data generated by a smart ring, worn on the finger, the user can be authenticated when making mobile payments or when knocking on a door (for access control). The proposed system can be deployed purely in software and does not require updates to existing payment terminals or infrastructure. We also demonstrate that smart ring data can authenticate smartwatch gestures, and \textit{vice versa}, allowing either device to act as an implicit second factor for the other. To validate the system, we conduct a user study (n=21) to collect inertial sensor data from users as they perform gestures, and we evaluate the system against an active impersonation attacker. Based on this data, we develop payment and access control authentication models for which we achieve EERs of 0.04 and 0.02, respectively.
\end{abstract}

\begin{IEEEkeywords}
wearable authentication, mobile payment, smart ring, smartwatch, tap gesture, knock gesture, authentication
\end{IEEEkeywords}

\section{Introduction}

Mobile payment systems (also known as tap-and-pay), such as Google Pay, have become pervasive in recent years. These systems enable the user to provision payment cards to a virtual wallet on a smartphone and then facilitate cashless and contactless payments with NFC-enabled point-of-sale terminals. The functionality of mobile payment systems has been extended to wearable devices, such as smartwatches. When paired with a smartphone, a smartwatch can access and store the same virtual wallet and make payments even when the smartphone is not present. However, the options for authenticating the user to a smartwatch are limited, given its size, and so the smartphone is still regarded as the primary device for the system. Recently, we are beginning to see commercial smart rings enter the market that offer payment capabilities, such as the Mastercard K-Ring\footnote{https://thepaymentring.com}, which have even fewer options for authentication. In order to protect payment transactions in these emerging systems, new authentication factors that operate on wearable devices are needed.

Given that wearable devices are often designed with continuous healthcare or fitness monitoring use-cases in mind, they tend to have an inertial measurement unit (IMU) consisting of (at least) an accelerometer and gyroscope. Works in behavioural biometrics have shown that inertial sensor data in smartphones and smartwatches can be used to infer gait or gestures. These systems require some initial calibration (\textit{i.e.}, a cumbersome enrolment phase), but can then continuously or discretely authenticate the user with reasonable effect.

In this work, we show that inertial sensor data generated by a smart ring can be used to authenticate the user when making certain gestures, such as tapping a payment terminal or knocking on a door. 

\textbf{Contributions.}
\begin{itemize}
\vspace{-\topsep}
\item We propose a novel, smart ring-based authentication system. Using only inertial sensor data, we show that a single tap gesture performed by a user while making a payment with a smart ring can implicitly authenticate the user. We also show that smart ring data can be used to authenticate a user making payments with a smartwatch, and \textit{vice versa}, allowing either device to act as an implicit second factor for the other.
\item We show that our approach can also be applied to an access control context, in which a single knock gesture against a door can explicitly authenticate the user.
\item We demonstrate that our authentication models are resistant against an active impersonation attacker.
\item We make the code and data required to reproduce our results available at http://github.com/jacksturgess.
\end{itemize}

\textbf{Paper Structure.} The rest of this paper is organised in the following way. Section \ref{sec:ObjectivesandAssumptions} states our objectives and details our system and threat models. Section \ref{sec:ExperimentalDesign} describes the design of our data collection user study. Section \ref{sec:Methods} explains the methods that we employ to process and classify our data. Section \ref{sec:Results} presents and analyses our results. Section \ref{sec:Discussion} discusses peripheral topics. Section \ref{sec:RelatedWork} compares our approach to related work. Section \ref{sec:LimitationsandFutureWork} considers the limitations of our approach. Section \ref{sec:Conclusion} concludes the paper.

\section{Objectives and Assumptions}\label{sec:ObjectivesandAssumptions}

\subsection{Design Goals}

In this paper, we investigate the use of a smart ring for authentication purposes, both implicit and explicit.

For implicit authentication, we select the context of mobile payments in which we consider the tap gesture made when the user taps an NFC-enabled device against a point-of-sale terminal to provide an implicit gesture biometric and we pose the following questions:
\begin{itemize}
\vspace{-\topsep}
\item can tap gestures made with a smart ring, as measured by the inertial sensors of the smart ring, be used to implicitly authenticate the user during a transaction?
\item can tap gestures made with a smart ring, as measured by the inertial sensors on a \textit{smartwatch} worn on the same arm, be used to implicitly authenticate the user (in cases where the smart ring does not have inertial sensors of its own) during a transaction?
\item can implicit authentication in a smartwatch system be improved by incorporating (the inertial sensor data of) a smart ring as a second factor?
\end{itemize}

For explicit authentication, we choose the act of knocking on a door. A knock gesture could be used implicitly for identification purposes or it could be performed solely for access control purposes, in which case it would be treated as an explicit authentication gesture. We pose the question: can (explicit) knock gestures, as measured by the inertial sensors of a smart ring, a smartwatch, or directly mounted on the door, be used to authenticate the user? We consider both unrhythmic knocks and user-chosen secret knocks.

\subsection{System Model}\label{sec:SystemModel}

For our payment model, we consider a system model in which the user is wearing both a smart ring and a smartwatch on the same arm and is using them to make NFC-enabled payments at point-of-sale terminals in a typical setting (\textit{e.g.}, in a shop). To make a payment, we assume that the user performs a \textit{tap gesture} by moving one of the devices towards the terminal until it is near enough to exchange data via NFC. The NFC contact point is when the payment provider would decide whether to approve the payment, so we assume that this marks the end of the tap gesture.

For our access control model, we consider a system model in which the user is wearing both a smart ring and a smartwatch on the same arm and is knocking on a door with that hand as a means to authenticate to an access control system. We assume that each \textit{knock gesture} is bounded by button-presses on one of the devices.

We assume that the devices have an accelerometer and gyroscope and that we have access to their data. We use data from the inertial sensors only. We assume that the user's biometric templates are stored securely on the wearable devices. When we combine data from multiple devices into a single sample for classification, as we do in some of our models, we assume that one device shares its inertial sensor data wirelessly with the other in a trusted fashion. We assume that the devices will be running a trusted app that is able to communicate its authentication decisions with the payment provider or access control system.

\begin{figure}[t!]
	\centering
  	\includegraphics[width=0.88\linewidth]{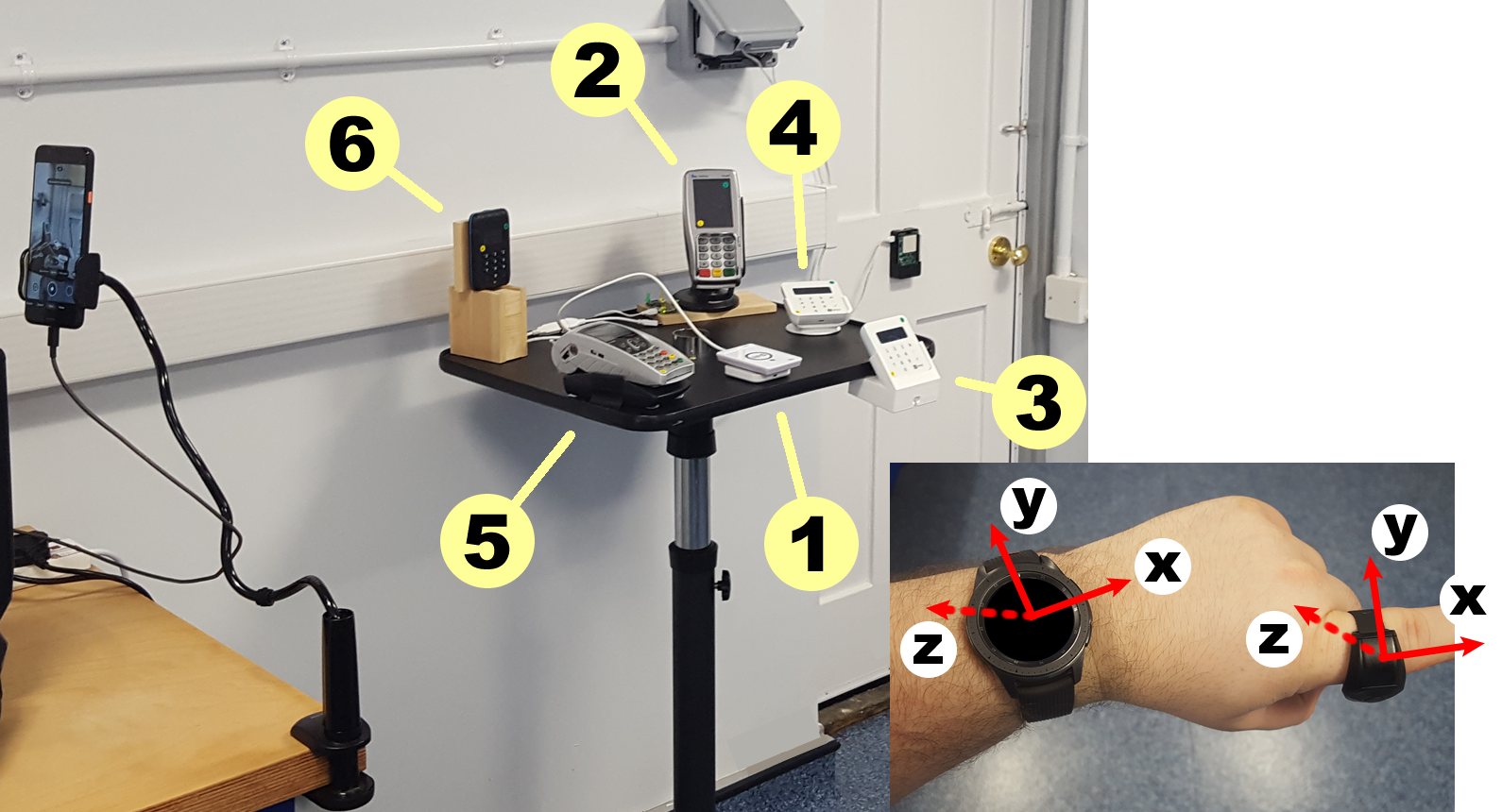}
   	\caption{The equipment used in our experiments: six fixed terminals (labelled, see Table \ref{tab:TerminalDetails} for details), an NFC reader (affixed to Terminal 1), a Raspberry Pi for timestamp collection, a Raspberry Pi attached to a door, and a fixed smartphone for video recording. Inset: our smart ring and smartwatch worn on the left arm with the sensor axes shown (the $z$-axis points upwards through the screen).}
   	\label{fig:ExperimentOverview}
\end{figure}

\begin{table}[t!]
	\centering
	\small
	\begin{tabular}{c|c|c|c}
		\toprule
		\textbf{Terminal} & \textbf{Height (cm)} & \textbf{Tilt ($^{\circ}$)} & \textbf{Distance (cm)} \\
		\midrule
		1 & 100 & 0 & 5 \\
		2 & 120 & 60 & 25 \\
		3 & 95 & 45 & -10 \\
		4 & 105 & 30 & 15 \\
		5 & 110 & 15 & 10 \\
		6 & 115 & 90 & 30 \\
		\midrule
		F & \multicolumn{3}{c}{picked up from centre of platform} \\
		\bottomrule
	\end{tabular}
	\caption{Details of the terminals used in our experiment; the indices match those labelled in Figure \ref{fig:ExperimentOverview} and `F' is the freestyle terminal. \textit{Height} is measured from the floor to the lowest point of the terminal; \textit{Tilt} is the inclination at the lowest point of the terminal; and \textit{Distance} is measured from the front of the stand to the point of the terminal that is closest to the user. Terminals 2 and 6 match terminals on self-service checkouts at supermarket chains.}
	\label{tab:TerminalDetails}
\end{table}

\begin{figure}[t!]
	\centering
  	\includegraphics[width=0.82\linewidth]{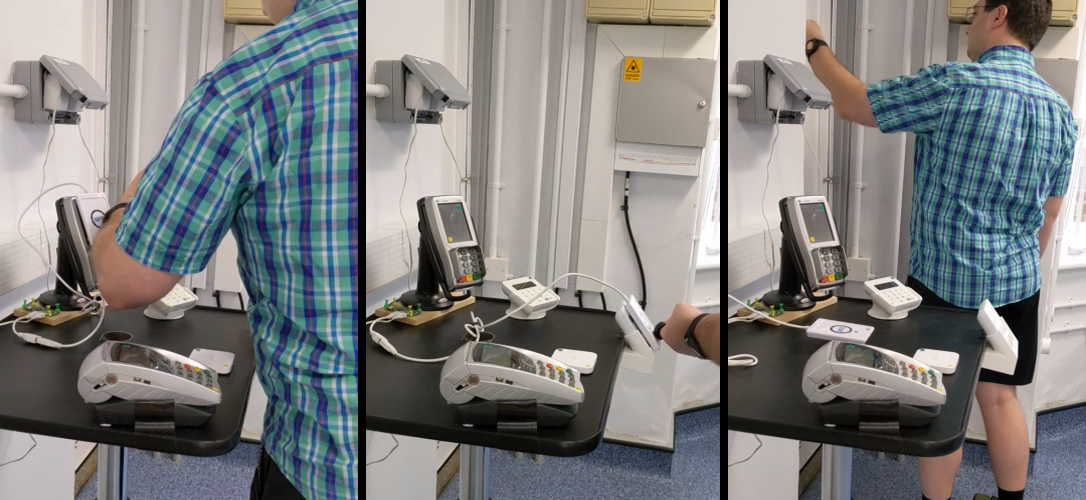}
   	\caption{The attacker's restricted view of a victim interacting with Terminal 2 (left), Terminal 3 (centre), and the door (right).}
   	\label{fig:ExperimentRecordings}
\end{figure}

\subsection{Threat Model}\label{sec:ThreatModel}

We consider an adversary that has possession of a legitimate user's smart ring (or smartwatch, or both, as appropriate), has unlocked it, is wearing it, and is attempting to use it to make a payment at a terminal or to gain entry to a locked door via an access control system. The adversary may have (maliciously) stolen the device(s) or (benignly) borrowed it. Our goal is to authenticate the legitimate user and to reject the adversary by using only inertial sensor data. We consider the following two types of attack:
\begin{itemize}
\vspace{-\topsep}
\item \textit{zero-effort attack}: for any given victim, all other users are considered to be passive, zero-effort attackers;
\item \textit{observation attack}: an active attacker who watches the victim perform gestures (\textit{e.g.}, via a hidden camera) and then attempts to mimic them.
\end{itemize}

In this work, we concentrate on the extent to which gesture biometrics can be used to defend against these attacks. We do not consider threats to other components in the system, tampering of devices or biometric templates, malware, or denial of service attacks.

\section{Experimental Design}\label{sec:ExperimentalDesign}

\subsection{Experiment Overview}

To evaluate the extent to which finger and wrist motion data can be used to authenticate users, and to compare the two, we designed and conducted a user study to collect data. Our experiments consisted of six point-of-sale terminals on an adjustable stand fixed at a height of 100 cm, an ACR122U NFC reader connected to a Raspberry Pi for timestamp collection, a second Raspberry Pi with an accelerometer and a gyroscope attached to a closed door, and a smartphone fixed in position for video recording. For our wearable devices, we used a smart ring and a smartwatch (detailed below), both commercial off-the-shelf devices and worn together on the same arm by the user. We collected motion data from the wearable devices and the door-mounted Raspberry Pi (with all clocks synchronised) as the user performed gestures. Figure \ref{fig:ExperimentOverview} shows our apparatus.

For our payment experiment, we affixed an NFC tag to the front of each wearable device and we affixed the NFC reader to the front of each point-of-sale terminal in turn. For each terminal, the user stood in front of the stand and performed tap gestures on the terminal using a wearable device, as if making mobile payments, with a short spacing delay between each tap gesture. The use of NFC tags and the NFC reader ensured consistent NFC communication between wearable devices and terminals. Each NFC tag stored the name of the wearable device to which it was affixed; each time an NFC contact point was made during a tap gesture, the Raspberry Pi captured its timestamp and the name of the triggering device. This was later used to segment the inertial sensor data collected from the wearable devices to retrieve the data for each tap gesture.

For our access control experiment, the user stood in front of the door and performed knock gestures against it using his device-wearing hand. The user pressed the button on the smart ring before and after each knock gesture to capture bounding timestamps. These timestamps were later used to segment the inertial sensor data collected from all devices to retrieve the data for each knock gesture.

\subsection{Point-of-Sale Terminals}

To emulate a real-world mobile payment setting as much as possible, we captured tap gestures using \textit{seven} terminals: six in fixed positions (as detailed in Table \ref{tab:TerminalDetails}) and one `freestyle'. For five of the six fixed terminals, we surveyed prominent supermarket and restaurant chains to find popular or standardised terminal positions (in terms of height, tilt angle, and distance from the user) and set our terminals to match common configurations. We set the other fixed terminal (Terminal 3) to match the position of the terminal on a train station barrier in the UK, which is widely standardised. For the freestyle terminal, the user picked up the NFC reader with his other hand and performed a tap gesture against it, returning it after each interaction, as if a shopkeeper had handed an unmounted terminal to a customer.

The six fixed terminals remained deactivated throughout the experiment because their functionality was not required. For consistent data collection, we affixed the NFC reader to each terminal when using it. As such, the terminals should be regarded only as fixtures that enforced positions, as well as a tool for immersing the user in a payment scenario.

\subsection{Wearable Devices and Sensor Modules}

For our smart ring, we used a Genki Wave\footnote{https://genkiinstruments.com/products/wave}. This ring is worn on the index finger and has a button that can be pressed with the thumb---we utilised this for timestamp collection in our access control experiment. We wrote a data collection script in Python using the open source library\footnote{https://github.com/genkiinstruments/genki-wave} provided by the developers to interface with the ring over Bluetooth LE.

For our smartwatch, we used a Samsung Galaxy Watch\footnote{https://www.samsung.com/uk/watches/galaxy-watch} running the Tizen 4.0 operating system. We built a data collection app and installed it on the smartwatch using the Tizen Studio IDE.

From each of these wearable devices, we collected timestamped data from four inertial sensors directly or derived from their MEMS sensors. The \textit{accelerometer} measures change in velocity. The \textit{gyroscope} measures angular velocity. The \textit{linear accelerometer} is derived from the accelerometer with the effects of gravity excluded. The \textit{gyroscope rotation vector} (GRV) is a fusion of sensor readings to compute the orientation of the device. We collected this data with sampling rates of 100 Hz for the smart ring (which we downsampled to 50 Hz), 50 Hz for the smartwatch, and 30 Hz for the door-mounted Raspberry Pi.

The inertial sensor axes are fixed relative to the frame of each device (as shown in Figure \ref{fig:ExperimentOverview}). Motion along the $x$-axis corresponds with arm extension or withdrawal; the $y$-axis, with side-to-side arm waving; and the $z$-axis, directly up- and downwards through the screen.

\subsection{Tap and Knock Gestures}\label{sec:ExperimentalDesignGestures}

To collect tap gestures for our payment experiment, we had each user perform sets of ten of each of the following tap gestures on each of the seven point-of-sale terminals:
\begin{itemize}
\vspace{-\topsep}
\item \textit{ring tap}: the user tapped the smart ring against the terminal and moved it around, if necessary, until NFC contact was made to simulate a \textit{ring}-based payment;
\item \textit{watch tap}: the user tapped the smartwatch against the terminal and moved it around, if necessary, until NFC contact was made to simulate a \textit{watch}-based payment.
\end{itemize}

To collect knock gestures for our access control experiment, we had each user perform sets of ten of each of the following knock gestures on the closed door:
\begin{itemize}
\vspace{-\topsep}
\item \textit{3-knock}: the user knocked on the door three times;
\item \textit{5-knock}: the user knocked on the door five times;
\item \textit{secret knock}: the user created a knock pattern consisting of between three and six knocks performed in a manner of his choosing (some users chose to knock to a rhythm, some included a twist or jolt of the wrist).
\end{itemize}

\subsection{User Study}\label{sec:ExperimentalDesignUserStudy}

To collect our data, we conducted a user study that was reviewed and approved by the relevant research ethics committee at our university. We recruited 21 participants, including staff, students, and members of the public. Each participant attended two data collection sessions on separate days. We collected 30 of each of the gestures detailed in Section \ref{sec:ExperimentalDesignGestures} (510 gestures in total) from each participant.

In each experiment, the participant was asked to stand facing the terminals or the door; aside from this, we did not prescribe any constraints on positioning as we wanted the user to interact comfortably as though acting in a real-world setting. The first three gestures of any type were performed in silence, to familiarise, then the reseacher engaged the participant in light conversation to simulate the distractions of a real-world environment. This sometimes elicited additional hand and body movements if the user gesticulated naturally.

\textbf{Impersonation.} To evaluate the robustness of our approach against an observation attacker, all 21 participants consented to having their gestures recorded and participated in an impersonation exercise. The first 3 were recorded as victims and the latter 18 impersonated them; then, at the end of the study, the first 3 were invited back to impersonate the other 18. This design also enabled us to compare the susceptibility of different victims and the skill of different attackers (also known as \textit{wolf and lamb analysis}; see Section \ref{sec:ResultsObservationAttack}). We recorded the following six gestures of each participant: smart ring and smartwatch tap gestures on Terminals 2 and 3, 5-knock gestures, and secret knock gestures. The camera was fixed in position, as if hidden, and so the amount of observable information was controlled; Figure \ref{fig:ExperimentRecordings} shows the attacker's view of the terminals and the different information observable for each type of gesture. For each of the six attacks, the attacking participant watched a short video of the victim performing the gesture three times and then made three attempts to mimic him. The attacker wore the wearable devices on the same arm as the victim during this exercise.

\textbf{User Statistics.} Of our 21 participants: 15 were male, 17 wore the devices on the left arm (the decision was led by the smartwatch; everyone who wore them on the right arm was female), 15 regularly wore a watch of some kind (7 of which wore a smartwatch), 13 had paid with a smartphone before, and 5 had paid with a smartwatch (\textit{i.e.}, 71\% of those who regularly wore one). 16 participants (76\%) remembered their secret knock, with an average of 4 days between their sessions (those who did not remember had an average of 4.2 days between sessions).

\section{Methods}\label{sec:Methods}

\subsection{Data Processing}

We collect time-series data from the four inertial sensors on our smart ring and smartwatch worn by the user and from the accelerometer attached to the door. Each accelerometer, gyroscope, or linear accelerometer sample is given in the form $(t, x, y, z)$ and represents the change in velocity or angular velocity along each axis at time $t$. Each GRV sample is given as a quaternion in the form $(t, x, y, z, w)$ and approximates the orientation of the device at time $t$.

We express a tap or knock gesture using a series of inertial sensor data samples within a time window. In our payment experiment, we retrieve the tap gestures for each user by segmenting 4-second blocks of sensor data using the NFC contact point timestamps as the endpoint of each window. We found that a 4-second maximum window size was sufficient to encapsulate the entirety of each tap gesture. To investigate optimum tap gesture parameters, we compare (in Section \ref{sec:Results}) the performances of gestures bounded by various window sizes and offsets, where the offset is the time between the NFC contact point and the end of the window. For an NFC contact point timestamp $T_0$, a window size $s$, and an offset $o$, we retrieve a tap gesture with start time $T_S$ and end time $T_E$, where $T_E=T_0-o$ and $T_S=T_E-s$. In our access control experiment, we retrieve the knock gestures for each user using the bounding timestamps captured when the user pressed the button on the ring before and after each gesture.

\subsection{Feature Extraction}

Whenever a gesture is retrieved, we apply a low pass filter to the data to reduce noise and then process the following five dimensions for each accelerometer, gyroscope, or linear accelerometer sample: the filtered $x$-, $y$-, and $z$-values, the energy of those filtered values, and the energy of the unfiltered (raw) values, where the energy of $\{x, y, z\}$ is given by $\sqrt{x^2+y^2+z^2}$. As GRV samples are expressed as quaternions, for those we process only the four filtered values (since the Euclidean norm of a quaternion is always 1). In total, we process each gesture in 19 dimensions.

For each gesture, we extract the following ten statistical features in each dimension: \textit{minimum}, \textit{maximum}, \textit{mean}, \textit{median}, \textit{standard deviation}, \textit{variance}, \textit{inter-quartile range}, \textit{kurtosis}, \textit{skewness}, and \textit{peak count}. We also calculate the \textit{mean} and \textit{maximum velocities} along each axis, the \textit{displacement} along each axis, and the \textit{Euclidean displacement} from each of its accelerometer, gyroscope, and linear accelerometer vectors, adding another 30 features. Ultimately, we reduce each gesture to a feature vector containing 220 members.

In previous work \cite{Sturgess2022-1, Sturgess2022-2}, we found this feature set to be ideal by starting with a larger set and pruning it down using normalised Gini importances to reject the least informative features, so we use it again here on similar grounds. This approach also yielded promising preliminary results in our access control model, so we used it for both models to allow for comparability and consistency throughout the paper.

In some of our models, we combine inertial sensor data from multiple devices. In these cases, the above features are extracted for each separate source and then concatenated together to form a linearly-larger feature vector.

\subsection{Classification}\label{sec:MethodsClassification}

We use random forest classifiers in each of our payment authentication and access control models. Similar works \cite{Acar2020, Sturgess2022-2} have shown that random forests are efficient, able to estimate the importance of features, and robust against noise. To balance relevance with learning time, we include 100 trees in each forest. To reduce the impact of random generation on our results, and to ensure that our results are fair and unselective, we train and test each classifier ten times with different forest randomisation seeds and average the outcomes.

\textbf{Terminal-agnostic Model.} To evaluate the zero-effort attacker, we train a set of classifiers that are user-dependent and terminal-agnostic. This means that a separate template (and decision threshold) is generated for each user and that, for each tap gesture under test, the tap gesture samples used to train the classifier came from tap gestures made only against other terminals (\textit{i.e.}, a leave-one-out approach). The user's tap gestures form the positive class and all other user's tap gestures form the negative class. As this is an authentication scenario, we ensure that the training data precedes the testing data by taking the tap gestures collected in users' first data collection session as training data (analogous to the enrolment phase, where the user template is created) and those collected in the second session as testing data (analogous to an authentication phase).

\textbf{Terminal-known Model.} To evaluate the observation attacker, we apply a similar design, except that we do not exclude tap gestures based on the terminal. We assume that, since the attacker has already observed the victim using the terminal in question, that the system has knowledge of the victim's tap gestures made against that terminal. We pair up every user as a victim with every other user as an attacker, one at a time, and train the classifiers, excluding all of the attacker's tap gestures. This enables us to generate the victim's decision threshold for that pairing, tuned to the EER (which we take as our baseline FAR, the \textit{base-FAR}), with no knowledge of the attacker, and then we test the attacker's impersonation samples against that tuned classifier to find his attack success rate (the \textit{observation-FAR}). We compare the two FARs to measure the success of the observation attack (see Sections \ref{sec:MethodsPerformanceMetrics} and \ref{sec:ResultsObservationAttack} for more details).

\textbf{Terminal-specific Model.} For investigative purposes, we also train a set of classifiers in which each is trained and tested on tap gestures performed only on a single terminal. This model enables us to measure the effectiveness of our approach if implemented on standardised, single-terminal systems, such as public transport systems.

\textbf{Access Control Model.} For knock gestures, to evaluate each of the attackers, we use a similar approach as with each respective payment authentication model above, except that we do not need to generalise the model over multiple terminals. We train separate models for each of the three knock gestures.

\subsection{Performance Metrics}\label{sec:MethodsPerformanceMetrics}

In each model, the \textit{true positives} is the number of times that the positive class (\textit{i.e.}, the legitimate user) is correctly accepted; the \textit{true negatives} is the number of times that the negative class (\textit{i.e.}, the adversary) is correctly rejected; the \textit{false positives} is the number of times that the negative class is wrongly accepted; and the \textit{false negatives} is the number of times that the positive class is wrongly rejected. The decision threshold, $\theta$, is the score at which the classifier chooses to assign to a sample the positive class rather than the negative. To tune a classifier, we adjust $\theta$ to modify the trade-off between security and usability; a larger $\theta$ is more resilient to false positives and thus favours security, a smaller $\theta$ favours usability.

To quantify the performance of our models, we find the optimum decision threshold such that the false acceptance rate (FAR) equals the false rejection rate (FRR); this point is called the equal error rate (EER). The FAR inversely indicates \textit{security}, by measuring the likelihood that the negative class will be wrongly accepted. The FRR inversely indicates \textit{usability}, by measuring the likelihood that the positive class will be wrongly rejected. The FAR and FRR are antagonistic insofar as setting $\theta$ to favour one will disfavour the other, so there will always be a point at which they cross over; the EER is a measure of system performance when consideration is balanced evenly between security and usability and is commonly used as a metric in authentication systems.

\begin{figure*}[t!]
	\vspace{-1em}
	\centering
	\begin{tabular}{ccc}
		\subfloat[ring tap; ring data]{
			\includegraphics[height=2.6cm]{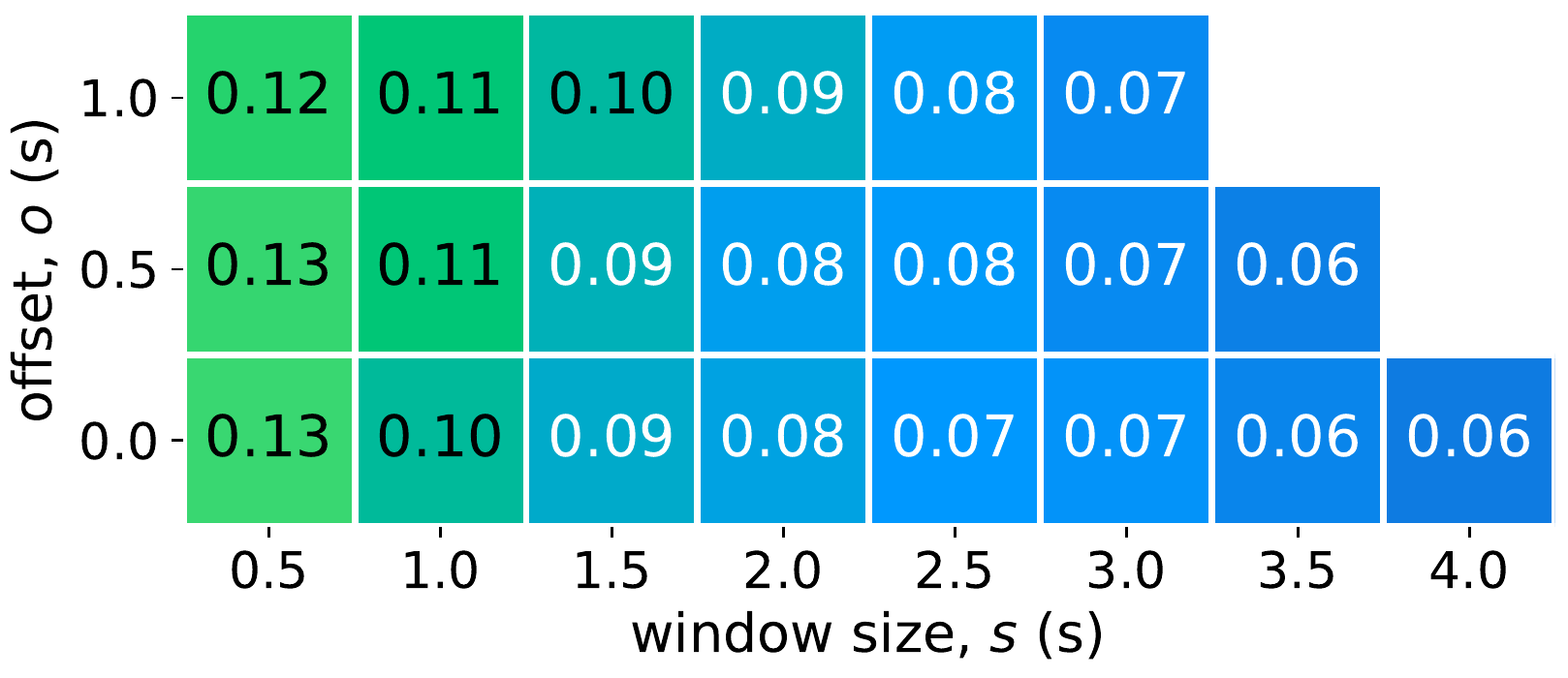}
			\label{fig:ResultsDownRingRing}
		} & 
		\subfloat[watch tap; ring data]{
			\includegraphics[height=2.6cm]{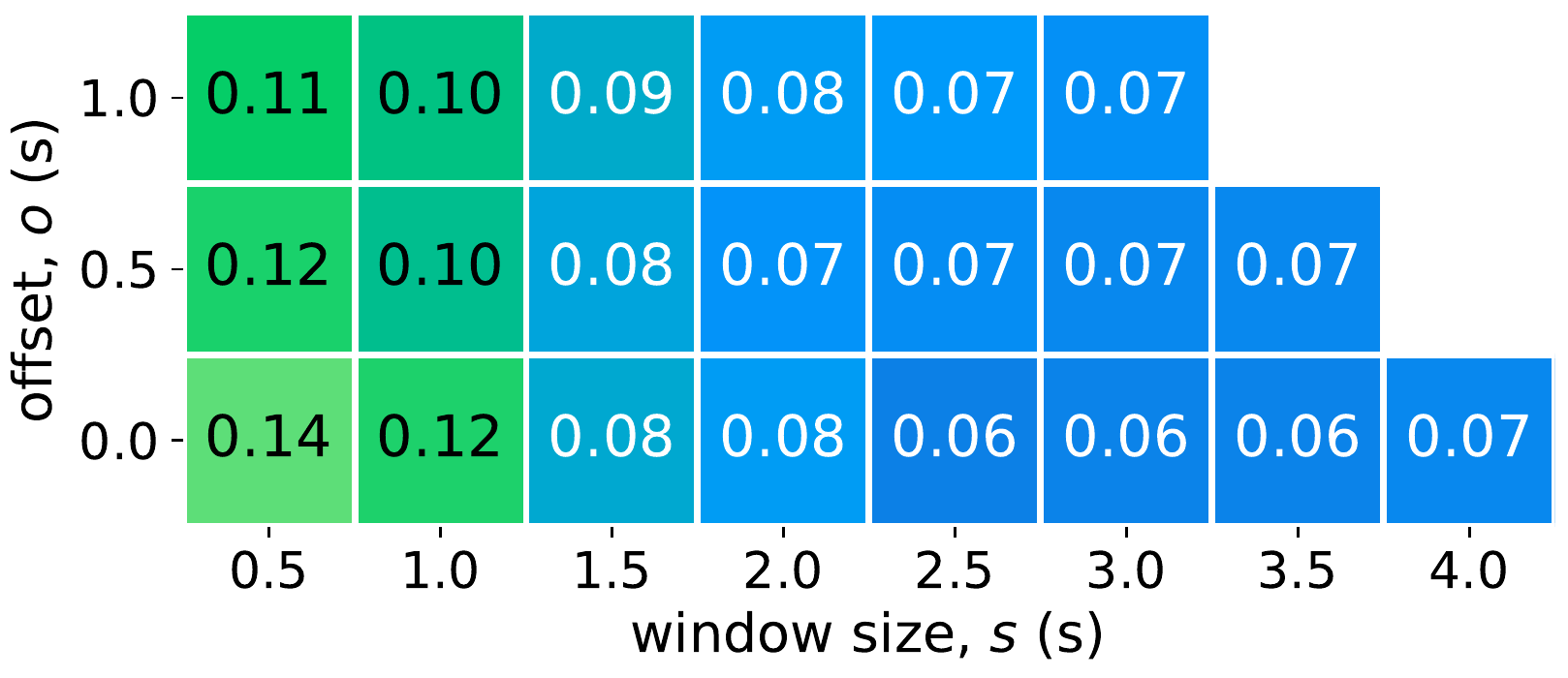}
			\label{fig:ResultsDownWatchRing}
		} & 
		\multirow{-3}[3]{*}{\subfloat{\includegraphics[height=6.0cm]{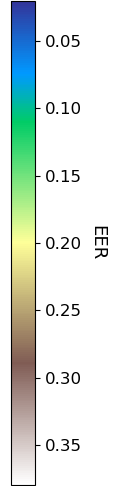}}} \\
		\setcounter{subfigure}{2}%
		\subfloat[ring tap; watch data]{
			\includegraphics[height=2.6cm]{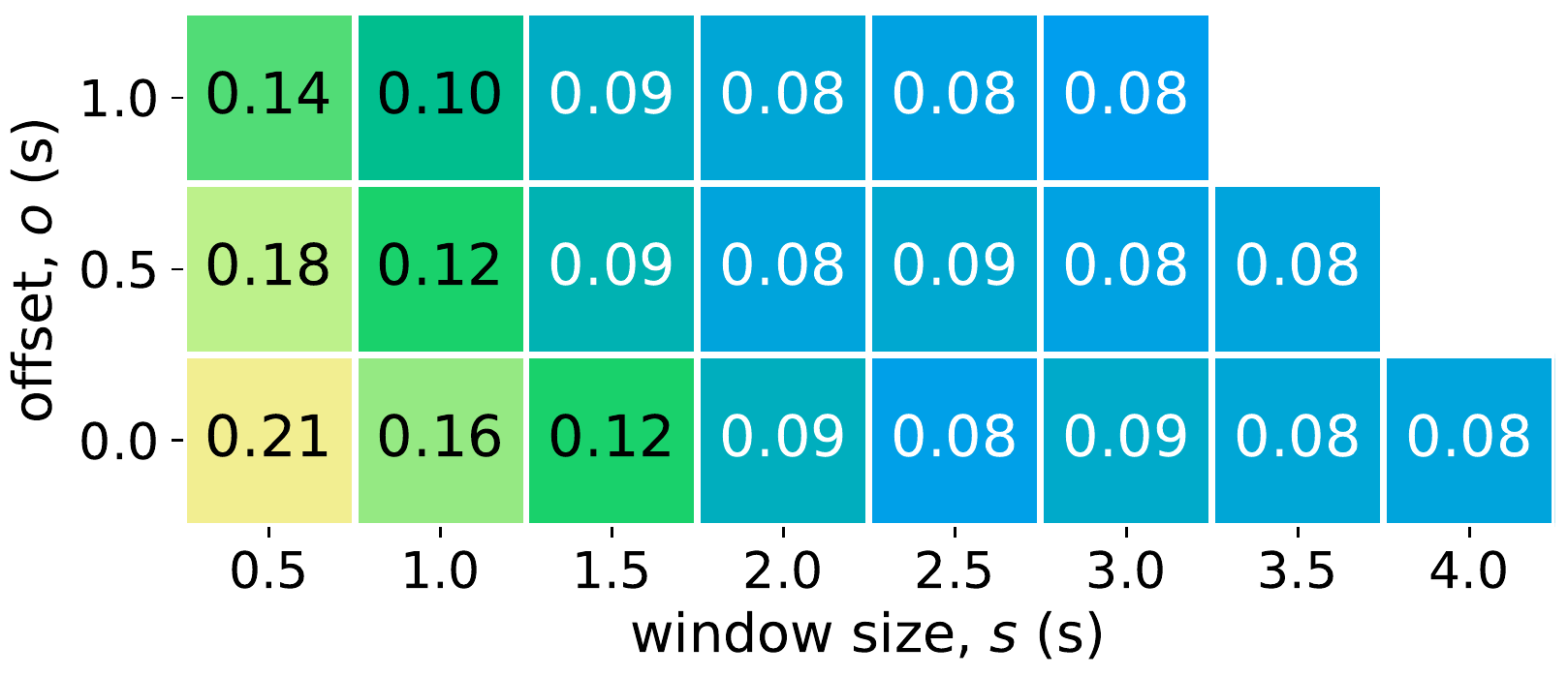}
			\label{fig:ResultsDownRingWatch}
		} & 
		\subfloat[watch tap; watch data]{
			\includegraphics[height=2.6cm]{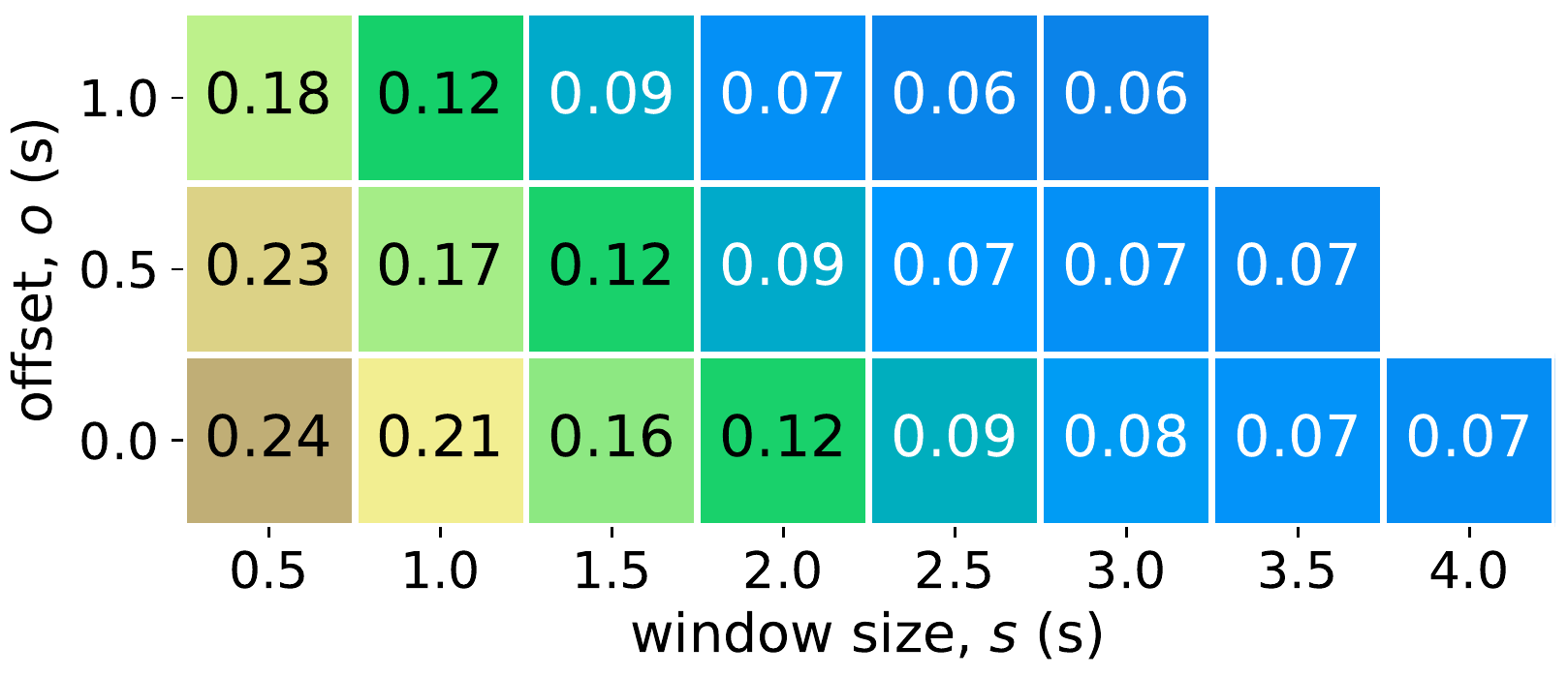}
			\label{fig:ResultsDownWatchWatch}
		} \\
		\subfloat[ring tap; combined data]{
			\includegraphics[height=2.6cm]{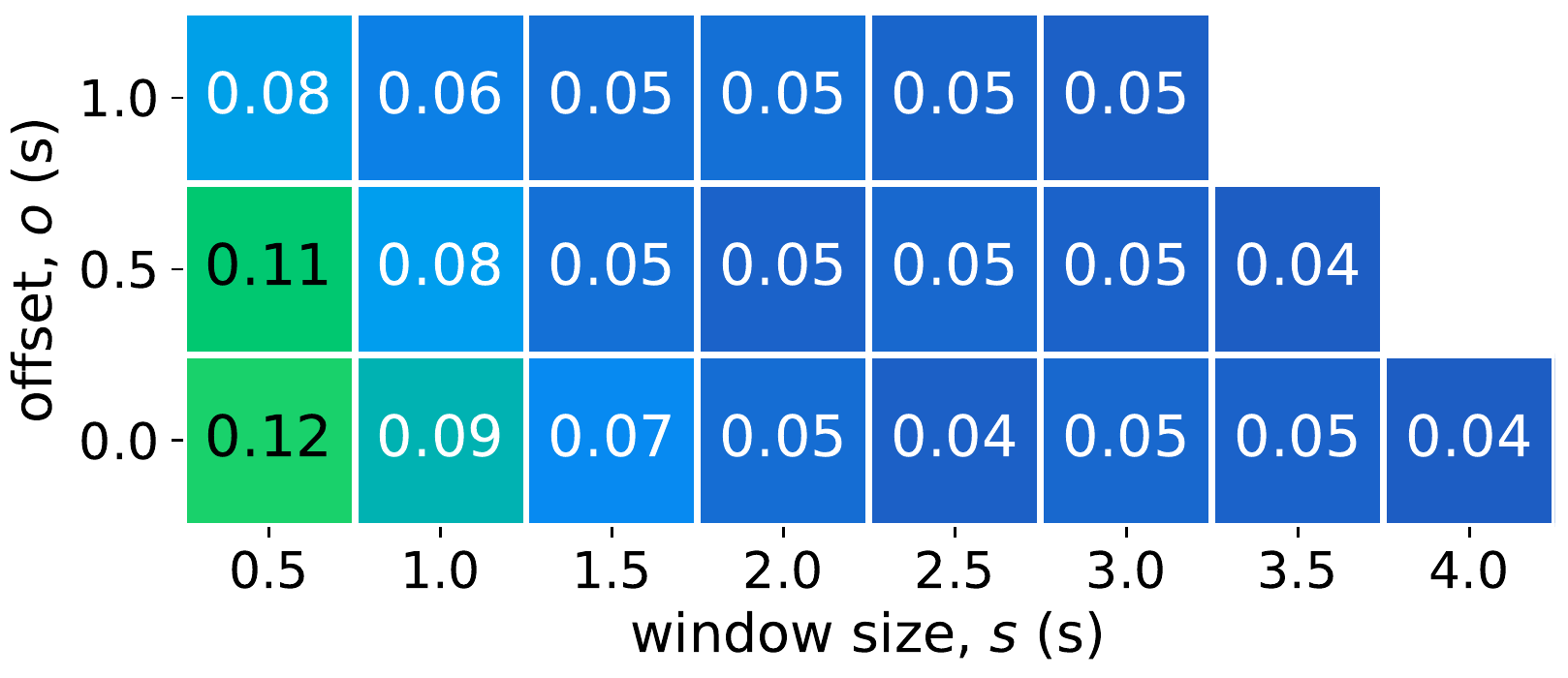}
			\label{fig:ResultsDownRingCombined}
		} & 
		\subfloat[watch tap; combined data]{
			\includegraphics[height=2.6cm]{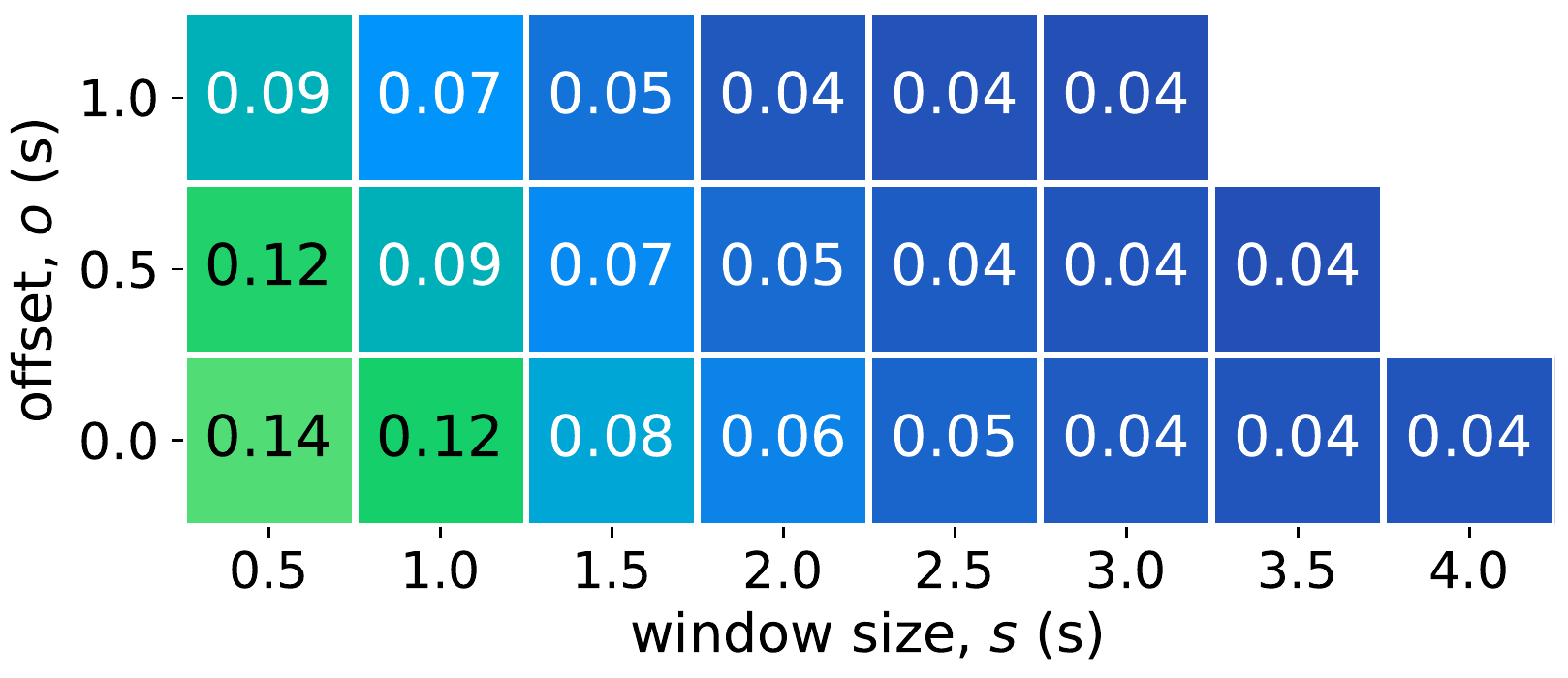}
			\label{fig:ResultsDownWatchCombined}
		} \\
	\end{tabular}
	\caption{Average EERs for our payment authentication models by window size and offset, for tap gestures made with the smart ring (left) and smartwatch (right), using data from the smart ring only (top), the smartwatch only (middle), and both combined (bottom). (This figure is based on downsampled smart ring data; Figure \ref{fig:ResultsUndown} in the Appendix is based on undownsampled smart ring data for comparison.)}
	\label{fig:ResultsDown}
\end{figure*}

\section{Results}\label{sec:Results}

\subsection{Zero-effort Attack}

\textbf{Tap Gestures.} Figure \ref{fig:ResultsDown} shows the average EERs for our \textit{terminal-agnostic} payment authentication models by window size and offset. With 21 users and 6 fixed terminals, each score is the average of scores from $21\times6\times10 = 1,260$ classifiers (see Section \ref{sec:MethodsClassification} for details).

For ring tap gestures, Figure \ref{fig:ResultsDownRingRing} shows that when using data from the smart ring only our model achieves EERs as low as 0.06. Figure \ref{fig:ResultsDownRingWatch} shows that when using data from the smartwatch only, we get 0.08. This suggests that a smartwatch could perform as a reasonable authenticator for a smart ring, in a scenario where the smart ring does not have any inertial sensors of its own (such as the K-Ring). If the data from both devices are combined, Figure \ref{fig:ResultsDownRingCombined} shows that we achieve EERs as low as 0.04. The optimum parameters for ring tap gestures, by considering EERs and favouring a smaller window size for usability, are $\{s=2.5, o=0\}$.

When we use data from the smartwatch only, we see a pattern radiating from the bottom left corner of the heatmap. This shows that, when training only on data where the tapping device is near to the terminal (small window, just before the NFC contact point is found), the movements of the wrist are not discriminative between users. The same is not true when using data from the smart ring only; this suggests that the finger remains active during that time, even for watch tap gestures. Figure \ref{fig:ResultsDownWatchCombined} shows that including a smart ring as an additional factor can improve the authentication of a user making watch-based payments (\textit{cf.} Figure \ref{fig:ResultsDownWatchWatch}).

\textbf{Knock Gestures.} Across all of the participants in our user study, the average 3-knock gesture lasted 2.81 seconds, the average 5-knock, 3.32 seconds, and the average secret knock, 3.78 seconds. Table \ref{tab:ResultsDoor} shows the average EERs for our access control models. Each score is the average of scores from $21 \times 10 = 210$ classifiers. Unexpectedly, our 3-knock model performed best, achieving an EER of 0.02 using data from the smartwatch only. This might be due to some participants sometimes miscounting the number of knocks when performing the 5-knock gesture, leading to messier data, whereas the 3-knock gestures were performed effortlessly and more consistently.

When we use data from the smartwatch only, we achieve the best results across all models, including those based on combined data. This suggests that wrist movements are a key discriminator in knocking, to such an extent that other factors act as pollutants. When we use data from the door only, we achieve the poorest results; the lower sampling rate of the door-mounted sensors may have had an impact, but the magnitude of the difference in results is likely explained by those sensors lacking knowledge of user movements.

\begin{table}[t!]
	\centering
	\small
	\setlength{\tabcolsep}{0.4em}
	\begin{tabular}{l c c c c}
		\toprule
		\textbf{Knock} & \\
		\textbf{Gesture} & \textbf{Door} & \textbf{Ring} & \textbf{Watch} & \textbf{Combined} \\
		\midrule
		3-knock & 0.17 & 0.06 & 0.02 & 0.03 \\
		5-knock & 0.21 & 0.12 & 0.04 & 0.05 \\
		secret knock & 0.19 & 0.09 & 0.05 & 0.05 \\
		\bottomrule
	\end{tabular}
	\caption{Average EERs for our access control models, using data from (i) the door-mounted sensors only, (ii) the smart ring only, (iii) the smartwatch only, and (iv) all three combined.}
	\label{tab:ResultsDoor}
\end{table}

\begin{figure*}[t!]
	\vspace{-1.2em}
	\centering
	\begin{tabular}{cccc}
		\hspace{-4em}
		\subfloat[ring tap; ring data]{
			\includegraphics[height=3.16cm]{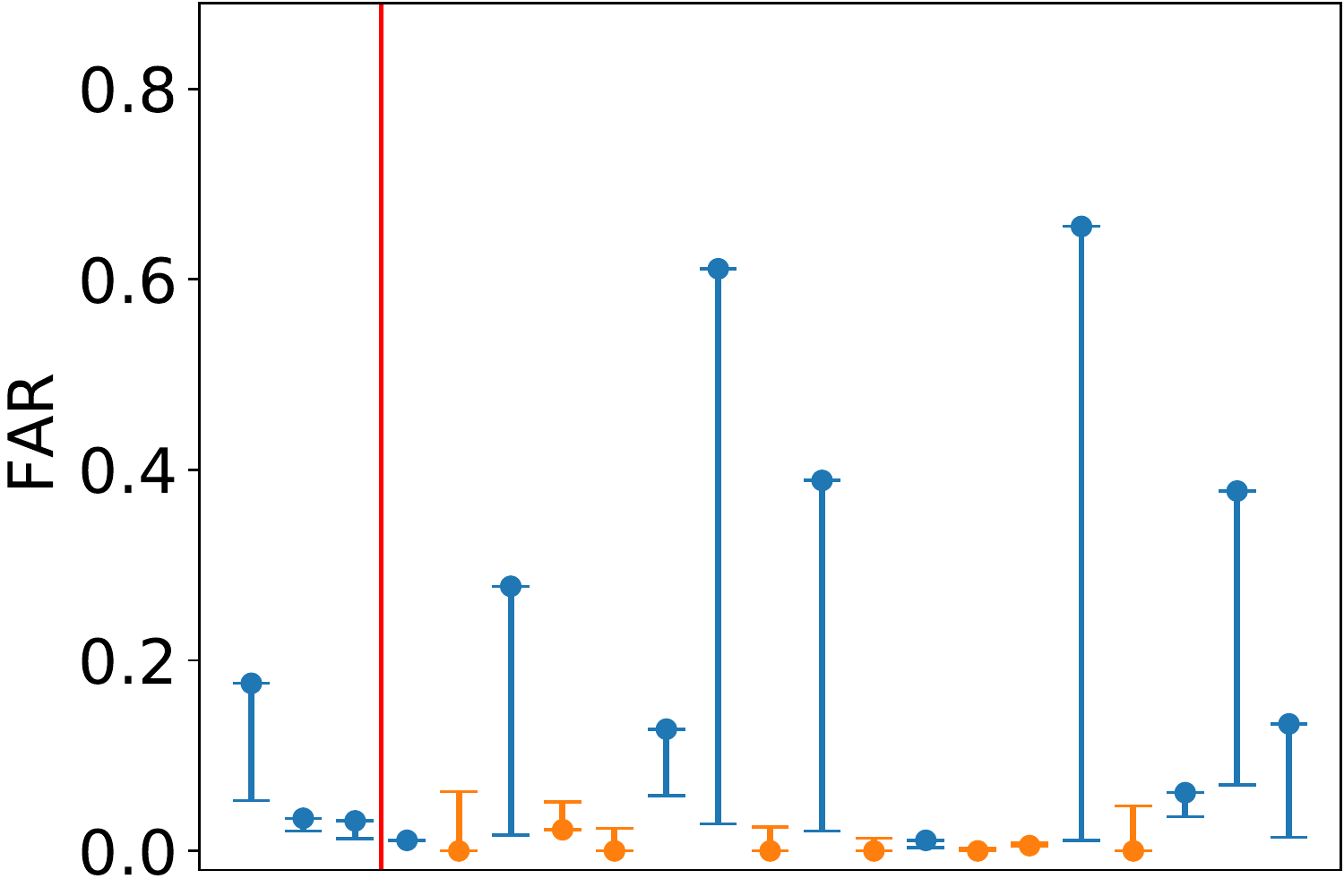}
			\label{fig:ResultsObservationRingRingVictim}
		} & 
		\hspace{-2em}
		\subfloat[ring tap; combined data]{
			\includegraphics[height=3.16cm]{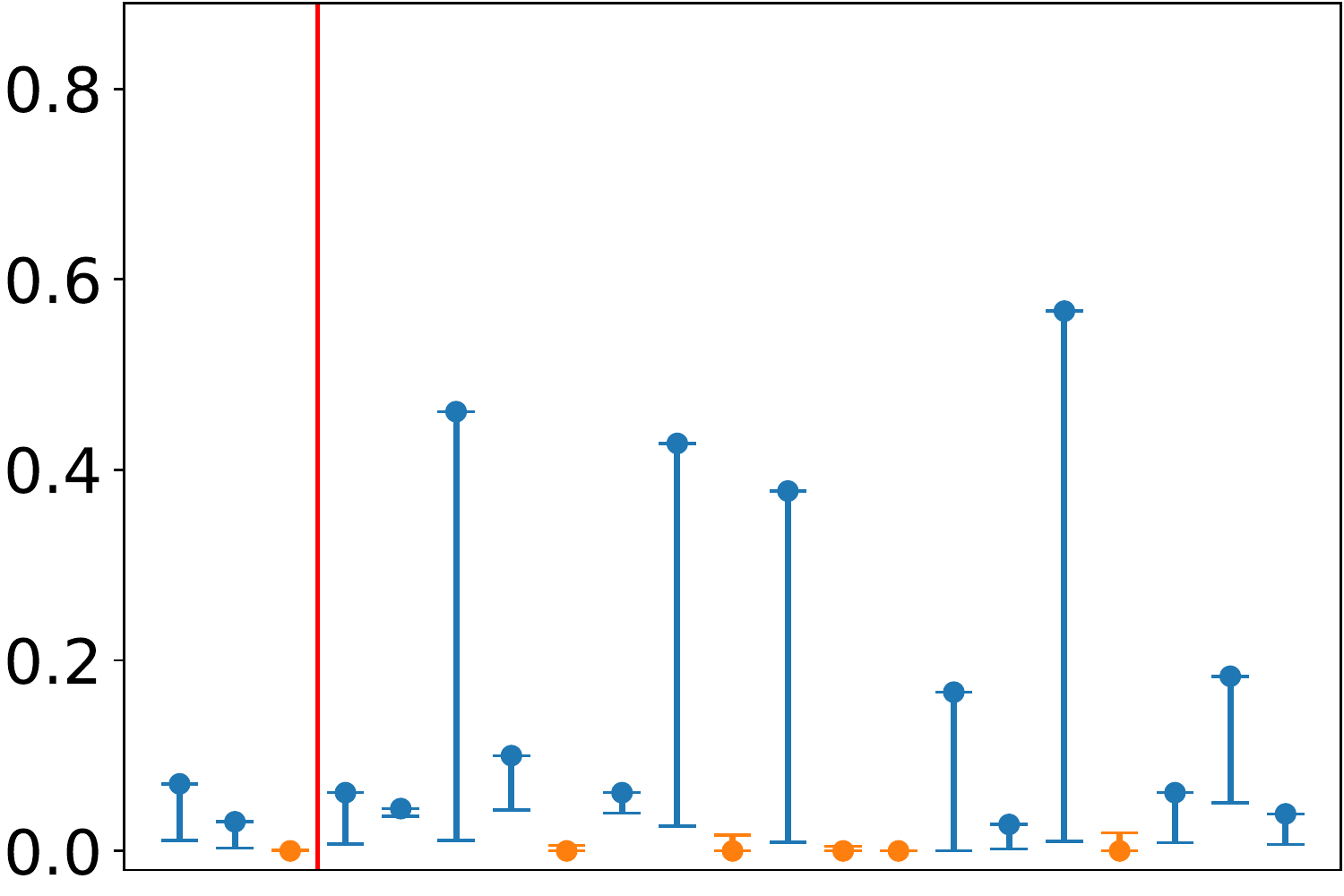}
			\label{fig:ResultsObservationRingCombinedVictim}
		} & 
		\hspace{-2em}
		\subfloat[5-knock; combined data]{
			\includegraphics[height=3.16cm]{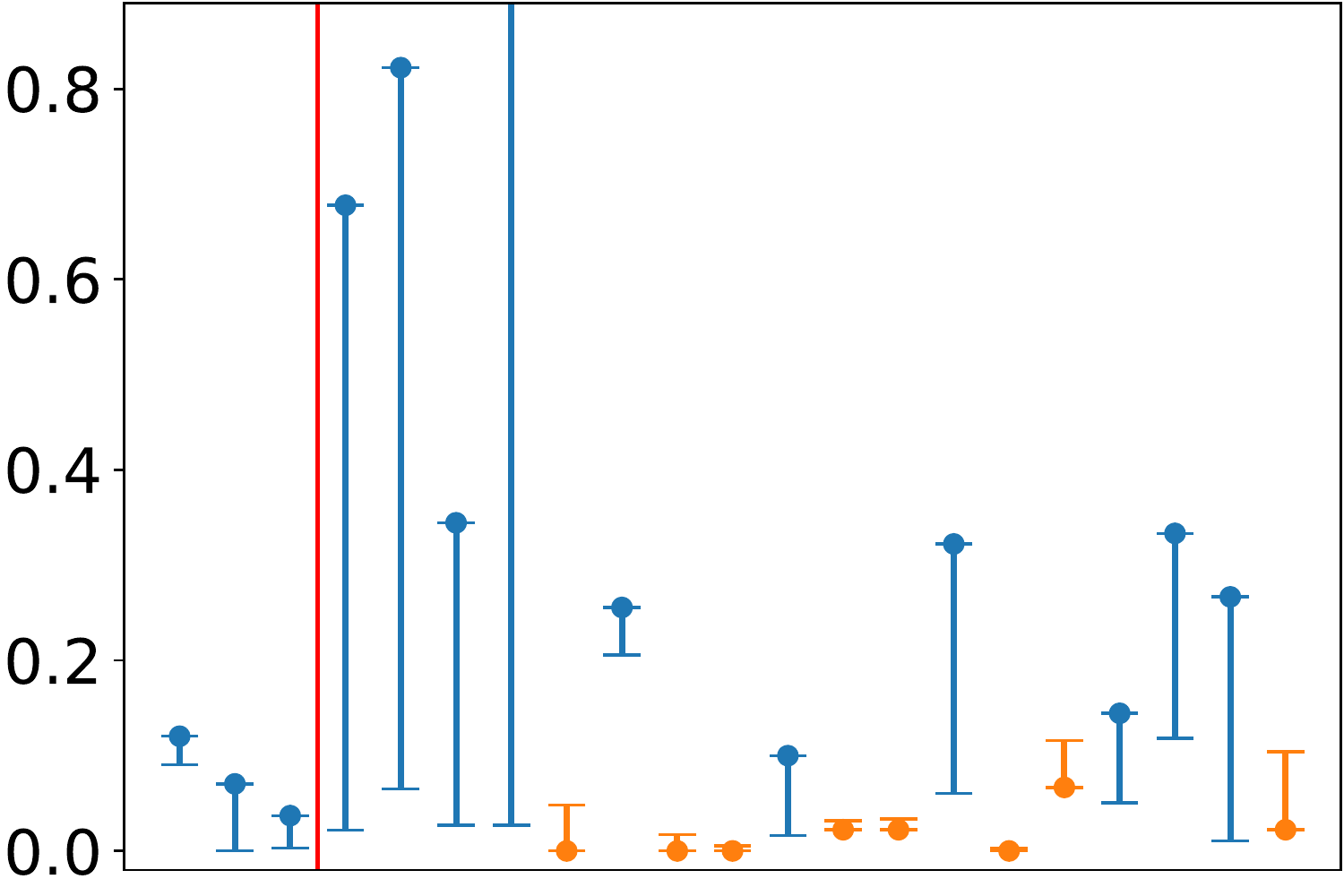}
			\label{fig:ResultsObservationNOC5CombinedVictim}
		} & 
		\hspace{-2em}
		\subfloat[secret knock; combined data]{
			\includegraphics[height=3.16cm]{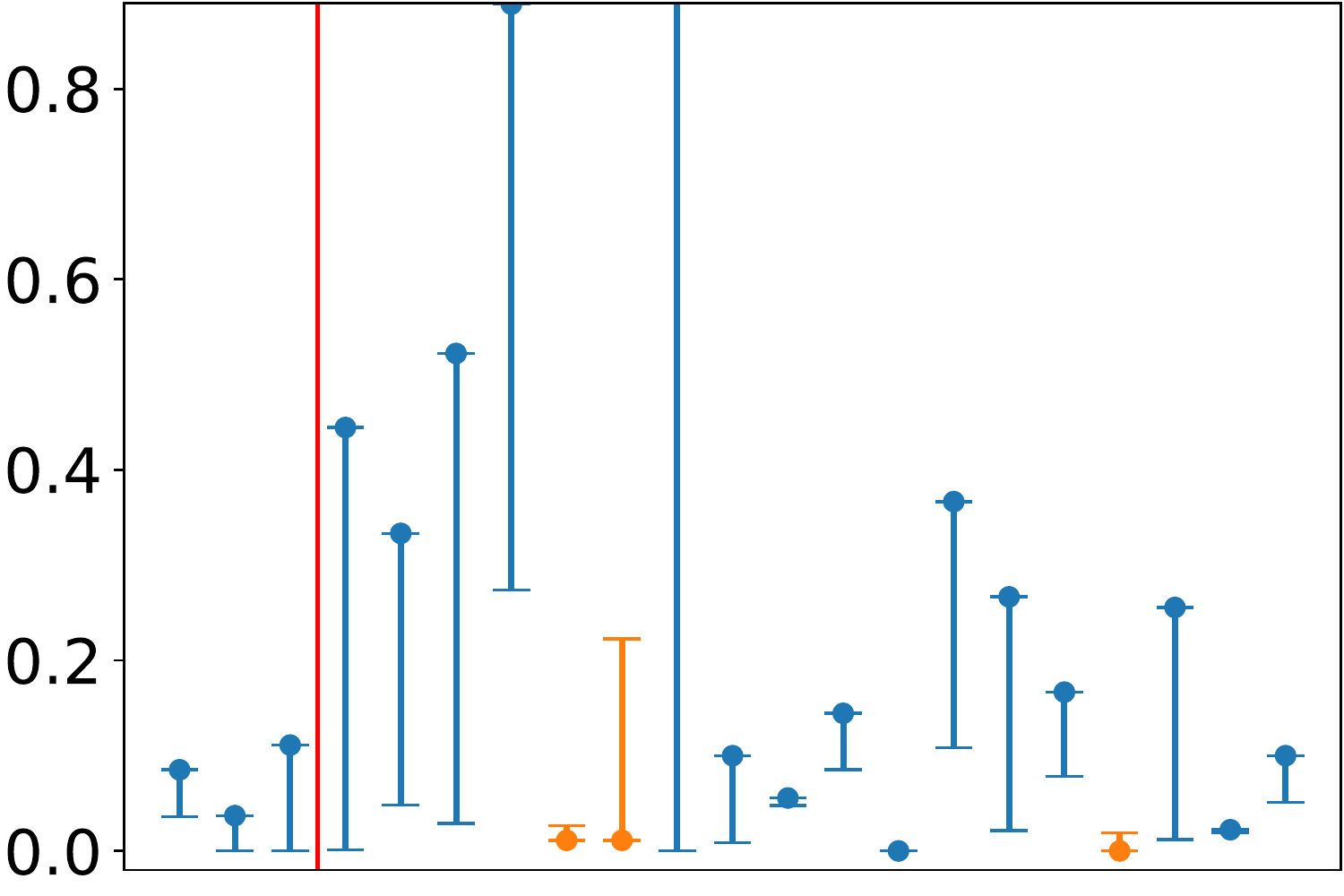}
			\label{fig:ResultsObservationNSECCombinedVictim}
		} \\
		\hspace{-4.5em}
		\subfloat[ring tap; ring data]{
			\includegraphics[height=1.62cm]{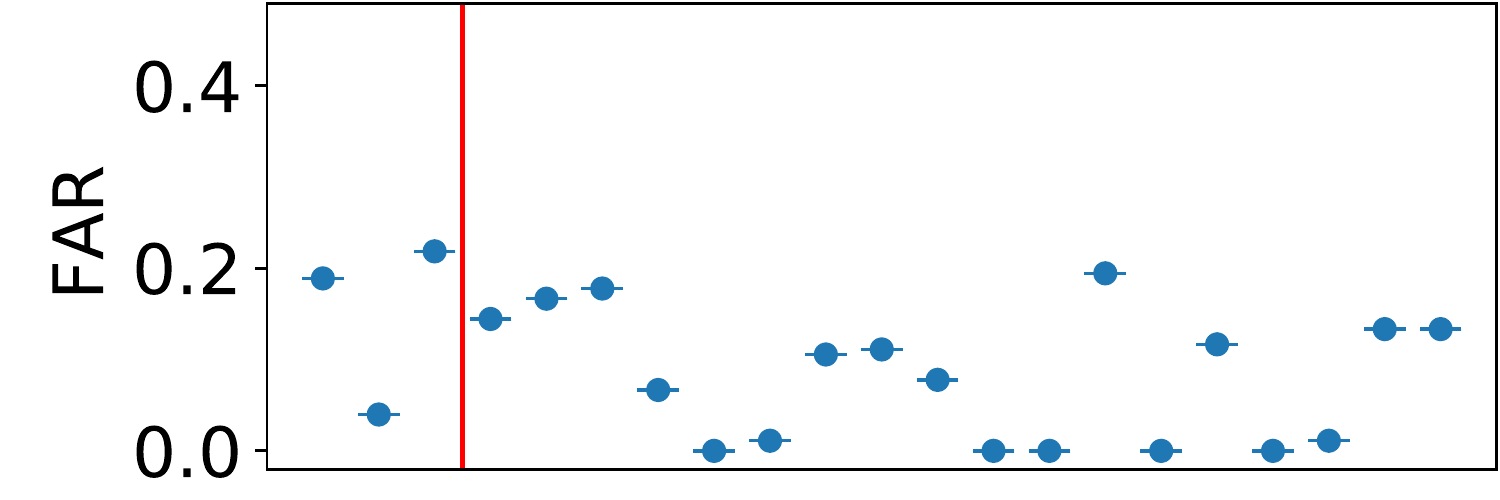}
			\label{fig:ResultsObservationRingRingAttacker}
		} & 
		\hspace{-2.2em}
		\subfloat[ring tap; combined data]{
			\includegraphics[height=1.62cm]{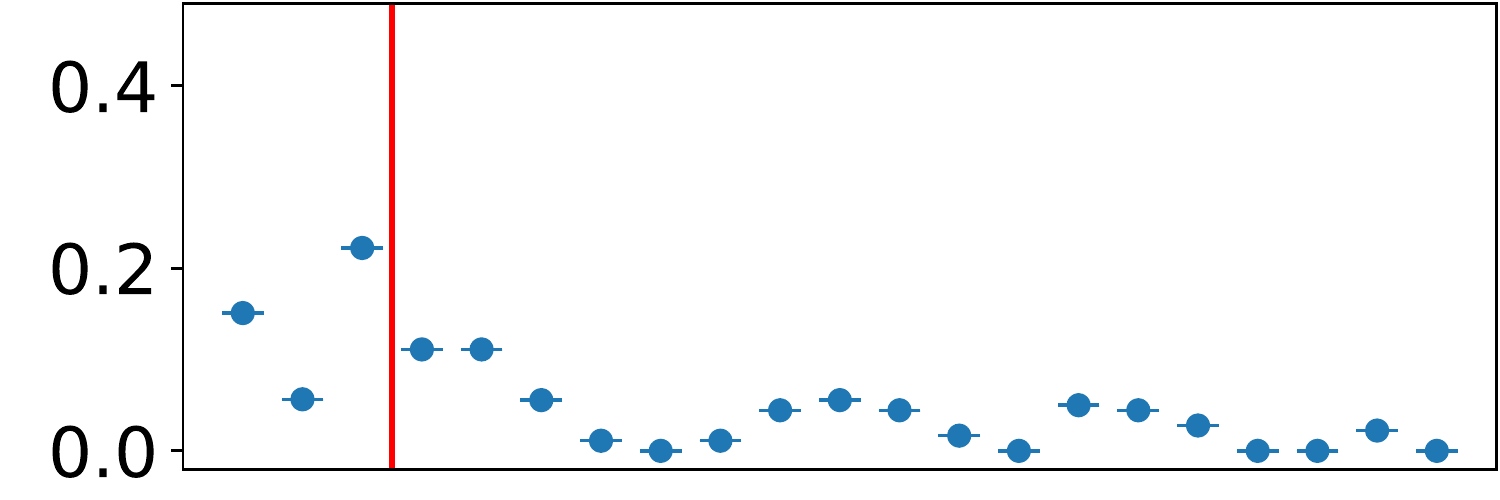}
			\label{fig:ResultsObservationRingCombinedAttacker}		
		} & 
		\hspace{-2.22em}
		\subfloat[5-knock; combined data]{
			\includegraphics[height=1.62cm]{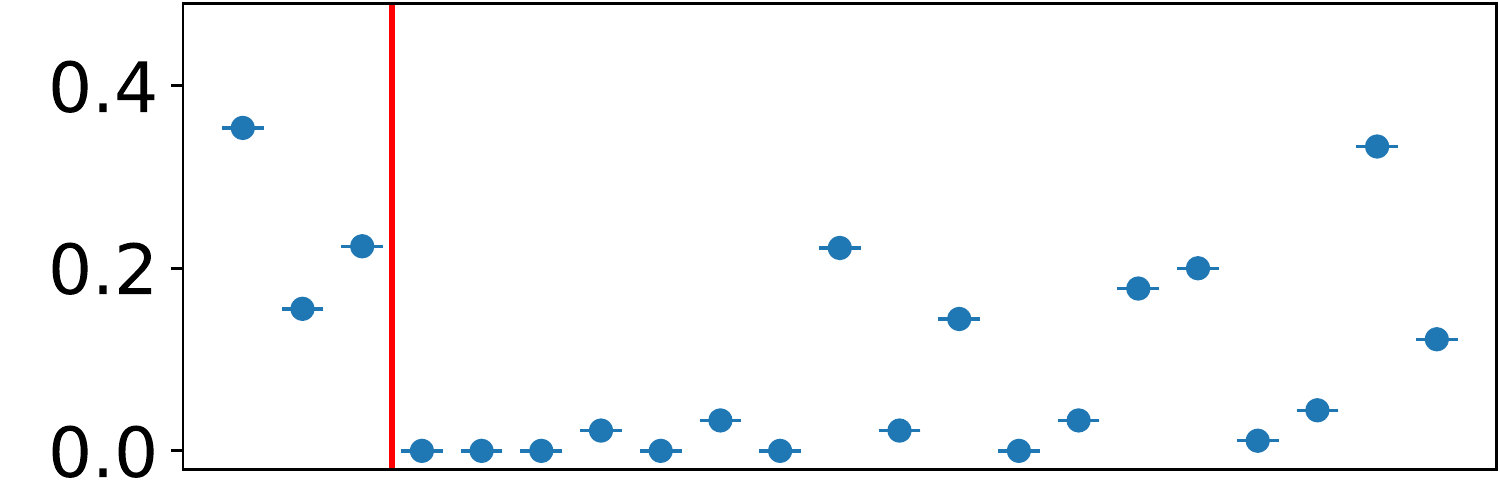}
			\label{fig:ResultsObservationNOC5CombinedAttacker}
		} & 
		\hspace{-2.24em}
		\subfloat[secret knock; combined data]{
			\includegraphics[height=1.62cm]{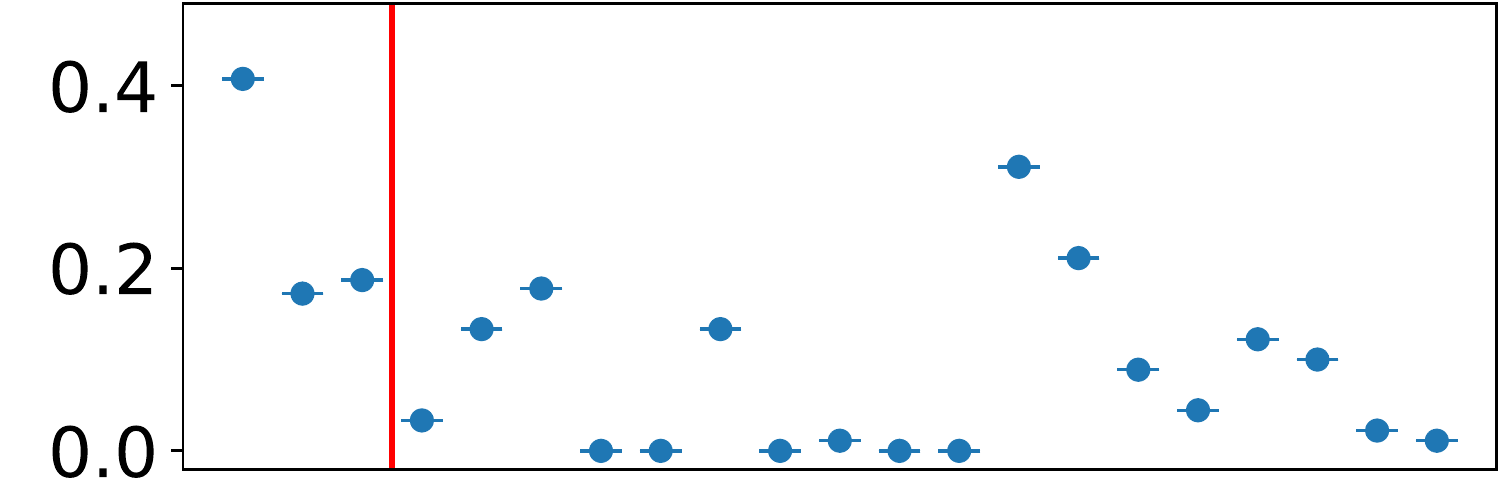}
			\label{fig:ResultsObservationNSECCombinedAttacker}
		}
	\end{tabular}
	\caption{Results of our observation attack against our \textit{terminal-known} payment authentication model in optimum window $\{s=2.5, o=0\}$ and our access control model. The top row shows for each user as a victim the average FAR of the user-specific base model (flat line) and the average FAR when attacked (circle); if the latter is greater, then the victim's line is coloured blue, otherwise it is orange. The bottom row shows for each user as an attacker the average FAR achieved when attempting to impersonate other users. The red line separates the first 3 users from the other 18, indicating the two groups of users that attacked each other (see Section \ref{sec:ExperimentalDesignUserStudy} for details).}
	\label{fig:ResultsObservation}
\end{figure*}

\subsection{Observation Attack}\label{sec:ResultsObservationAttack}

Figure \ref{fig:ResultsObservation} shows the results of our observation attack against our \textit{terminal-known} payment authentication model and our access control model. Each of the first 3 users was the victim of 1,080 ring tap impersonation attempts ($18 \textrm{ attackers} \times \textrm{ring tap gestures on 2 terminals} \times 3 \textrm{ attempts at each gesture} \times 10$), 540 5-knock attempts, and 540 secret knock attempts; each of the other 18 users, separated in the figures by a red line, was the victim of 180 (3 attackers), 90, and 90 attempts, respectively.

For ring tap gestures, the base model achieves average EERs (base-FARs) of 0.03 when using data from the smart ring only and of 0.01 when using data from the smart ring and smartwatch combined; when attacked, the average success rates (observation-FARs) are 0.06 and 0.05, respectively (and we see similar results for watch tap gestures). Figures \ref{fig:ResultsObservationRingRingVictim} and \ref{fig:ResultsObservationRingCombinedVictim} show that a small number of our users are \textit{lambs} (\textit{i.e.}, users that are especially susceptible to impersonation), where their observation-FAR is significantly larger than their base-FAR. Figure \ref{fig:ResultsObservationRingCombinedVictim} shows that the addition of the smartwatch data helps to reduce the largest FAR deltas, and the overall average observation-FAR, but also opens a new vector that increases the susceptibility of some users. Figures \ref{fig:ResultsObservationRingRingAttacker} and \ref{fig:ResultsObservationRingCombinedAttacker} show that none of our users are \textit{wolves} (\textit{i.e.}, users that are especially skilled at impersonation); when using data from the smart ring only, some attempts got lucky against a random spattering of users, but when the smartwatch data was combined, the success rate dropped. This is an important result, as it suggests that our system provides resistance against wolves, reducing the likelihood that an attacker could predictably impersonate a given victim (and so, in the wider system model, this may act as a deterrent).

For knock gestures, when using data from the door-mounted sensors, smart ring, and smartwatch combined, we have average base-FARs of 0.05 for both the 5-knock and secret knock gestures and average observation-FARs of 0.08 and 0.09, respectively. Figures \ref{fig:ResultsObservationNOC5CombinedVictim} and \ref{fig:ResultsObservationNSECCombinedVictim} show that we have a number of lambs, this time with greater FAR deltas. Knock gestures are notably weak against impersonation if they are loud and slow. For the secret knock, the fourth and sixth users after the red line have high base-FARs, because those users chose common gesture fragments in their secret knocks. For the former (and the seventh), the gesture was loud and slow, as evidenced by the huge observation-FAR. For the latter, curiously, the observation-FAR is far lower than the base-FAR, suggesting that the gesture was difficult to mimic intentionally despite having commonality with other gestures. This gesture contained three fast knocks in the middle, which captured the attention of attackers only for them not to match the surrounding knocks. Figures \ref{fig:ResultsObservationNOC5CombinedAttacker} and \ref{fig:ResultsObservationNSECCombinedAttacker} show that attentive attackers can achieve good attack success rates, but this is due to the lambs being more vulnerable rather than wolf-like behaviour.

\begin{table*}
	\vspace{-1em}
	\centering
	\small
	\begin{tabular}{cc}
		\hspace{-2em}
		\subfloat[payment model in $\{s=2.5, o=0\}$, ring tap gesture]{
			\setlength{\tabcolsep}{0.4em}
			\begin{tabular}{cc cc cc}
       			\toprule
    	   		\multicolumn{2}{c}{\textbf{Ring}} & \multicolumn{2}{c}{\textbf{Watch}} & \multicolumn{2}{c}{\textbf{Combined}} \\\cmidrule(lr){1-2}\cmidrule(lr){3-4}\cmidrule(lr){5-6}
       			Feature & \# & Feature & \# & Feature & \# \\
   	    		\midrule
    	   		\texttt{GRV-y-med} & 309 & \texttt{Acc-x-min} & 297 & \texttt{w-Acc-x-min} & 185 \\
	       		\texttt{GRV-y-mean} & 269 & \texttt{Acc-y-max} & 208 & \texttt{w-Acc-y-max} & 151 \\
       			\texttt{GRV-x-mean} & 239 & \texttt{Acc-x-max} & 208 & \texttt{r-Gyr-z-mean} & 145 \\
   	    		\texttt{GRV-x-med} & 223 & \texttt{Acc-z-max} & 189 & \texttt{r-Acc-x-velomax} & 141 \\
    	   		\texttt{GRV-x-max} & 222 & \texttt{Gyr-z-velomean} & 138 & \texttt{r-GRV-x-max} & 135 \\
	       		\texttt{GRV-y-max} & 217 & \texttt{Gyr-z-mean} & 126 & \texttt{r-Acc-x-mean} & 128 \\
        		\texttt{Gyr-z-mean} & 192 & \texttt{Gyr-z-min} & 124 & \texttt{r-GRV-y-mean} & 122 \\
   	    		\texttt{Acc-x-mean} & 162 & \texttt{Gyr-z-disp} & 119 & \texttt{r-Acc-x-med} & 122 \\
    	   		\texttt{GRV-z-max} & 157 & \texttt{Acc-x-mean} & 117 & \texttt{r-GRV-x-med} & 116 \\
	       		\texttt{Acc-x-velomax} & 155 & \texttt{Acc-x-velomax} & 116 & \texttt{r-GRV-y-med} & 116 \\
        		\bottomrule
			\end{tabular}
			\label{tab:ResultsFeaturesPayment}
		} & 
		\hspace{-2em}
		\subfloat[access control model, 3-knock gesture]{
			\setlength{\tabcolsep}{0.4em}
			\begin{tabular}{cc cc}
       			\toprule
    	   		\multicolumn{2}{c}{\textbf{Ring}} & \multicolumn{2}{c}{\textbf{Combined}} \\\cmidrule(lr){1-2}\cmidrule(lr){3-4}
       			Feature & \# & Feature & \# \\
   	    		\midrule
    	   		\texttt{GRV-w-med} & 24 & \texttt{d-Acc-y-med} & 22 \\
	       		\texttt{Acc-y-disp} & 22 & \texttt{w-Acc-x-max} & 20 \\
       			\texttt{GRV-y-max} & 20 & \texttt{w-Gyr-z-var} & 17 \\
   	    		\texttt{GRV-w-mean} & 19 & \texttt{w-Gyr-y-med} & 17 \\
    	   		\texttt{GRV-z-med} & 19 & \texttt{w-Gyr-z-velomax} & 15 \\
	       		\texttt{Acc-z-mean} & 18 & \texttt{w-Gyr-z-stdev} & 15 \\
        		\texttt{Acc-z-med} & 18 & \texttt{w-LAc-z-disp} & 15 \\
   	    		\texttt{Gyr-y-velomax} & 18 & \texttt{w-GRV-y-iqr} & 15 \\
    	   		\texttt{GRV-x-mean} & 18 & \texttt{w-GRV-y-med} & 14 \\
	       		\texttt{LAc-x-disp} & 17 & \texttt{w-Gyr-y-min} & 12 \\
	       		\bottomrule
			\end{tabular}
			\label{tab:ResultsFeaturesDoor}
		} \\
	\end{tabular}
	\caption{Modal top-five features by Gini importance summed over classifiers for our payment authentication and access control models, using data from (i) the smart ring only, (ii) the smartwatch only, and (iii) both combined for the former and from (i) the smart ring only and (ii) the door-mounted sensors, smart ring, and smartwatch combined for the latter. Features are given in the format sensor-axis-statistic; for combined models, the leading character indicates the device to which the sensor belongs (door, ring, or watch).}
	\label{tab:ResultsFeatures}
\end{table*}

\subsection{Feature Informativeness}

To investigate which features are most informative to our models, we sum the top five features, sorted by Gini importance, of each classifier. (Note that, \textit{w.r.t.} the counts, there are six times more classifiers for the payment models.)

For ring tap gestures, Table \ref{tab:ResultsFeaturesPayment} shows that our models favour GRV-derived features when using data from the smart ring only, but accelerometer- and gyroscope-derived features when using data from the smartwatch only (similar to the features favoured in \cite{Sturgess2022-2}), indicating that the finger moves to a position faster and remains in a position longer than the wrist, whose movements are smoother. The combined model, which achieved stronger results than the others as we saw in Figure \ref{fig:ResultsDown}, had twice as many members in its feature vector and ended up favouring a similar set, echoing the relative importance of these features.

For 3-knock gestures, Table \ref{tab:ResultsFeaturesDoor} shows the dominance of features derived from the $y$- and $z$-axes of the wearable devices, which is to be expected as these measure the sideways and forward movements of the hand, respectively. Aside from these, two notable exceptions show greater importance: the $x$-axis of the smartwatch yields a single important feature, representing the maximum acceleration of the arm as it extends towards the door initially, and the median impact experienced by the door-mounted accelerometer, indicating that each user struck the door with consistent force.

\subsection{Sensor Selection}

We collected motion data from all of the inertial sensors available on our devices. Some devices are more limited in their offering---the accelerometer is the commonest sensor, as it is the smallest and cheapest. To assess the feasibility of our approach on devices with fewer sensors, we trained a set of sensor-specific models in which each classifier is trained and tested on data from a subset of sensors.

For models that use data from the smart ring only, we see distinctly poorer results whenever we remove the GRV sensor. With just an accelerometer, we see an increase of approximately 0.04 in every EER. Conversely, for models that use data from the smartwatch only, Figure \ref{fig:ResultsDownAccGyr} in the Appendix shows that we achieve better results when using only the accelerometer and gyroscope; this suggests that the other sensors pollute the smartwatch classifiers, echoing similar findings in related work \cite{Sturgess2022-2}. The combined models remain roughly unchanged, favouring ring features when the GRV is included and watch features when not.

\subsection{Terminal Positions}

We collected tap gestures performed against a range of terminals. Some systems, such as public transport systems, have highly standardised terminals (\textit{i.e.}, dedicated terminals that can be found set at the same position in many instances). To compare the effectiveness of our approach in a general setting against a standardised setting, we trained a set of terminal-specific models in which each classifier is trained and tested on data from a single terminal.

Table \ref{tab:ResultsPositions} shows the average EERs for our \textit{terminal-specific} payment authentication models when trained and tested on tap gestures from a single terminal. In general, we gain little improvement from restricting our system to a single terminal. We find that user comfort has a beneficial impact on authentication results. Terminals 5 and 6 show slightly improved results and these were the most comfortably positioned terminals for the majority of participants, who wore the devices on their left arm (indeed, if we reconstruct our models using data only from those users wearing the devices on their left arm, the relative gains are greater still). Likewise, the freestyle terminal shows improved results for watch tap gestures, as the watch was more awkward to tap than the ring and this terminal accommodated smoother movements when using it---however, it had the opposite effect for ring tap gestures, perhaps because most users tilted and moved the terminal a shorter distance when interacting with it with the ring than with the watch, eliciting a simpler punch gesture.

\begin{table}[t!]
	\vspace{-1em}
	\centering
	\small
	\begin{tabular}{cc}
		\hspace{-2em}
		\subfloat[ring tap gesture]{
			\setlength{\tabcolsep}{0.4em}
			\begin{tabular}{c c c}
				\toprule
				\textbf{Terminal} & \textbf{Ring} & \textbf{Combined} \\
				\midrule
				1 & 0.07 & 0.04 \\
				2 & 0.09 & 0.07 \\
				3 & 0.09 & 0.07 \\
				4 & 0.08 & 0.03 \\
				5 & 0.06 & 0.02 \\
				6 & 0.07 & 0.03 \\
				F & 0.10 & 0.09 \\
				\midrule
				agnostic & 0.07 & 0.04 \\
				\bottomrule
			\end{tabular}
			\label{tab:ResultsPositionsRing}
		} & 
		\hspace{-2em}
		\subfloat[watch tap gesture]{
			\setlength{\tabcolsep}{0.4em}
			\begin{tabular}{c c c}
				\toprule
				\textbf{Terminal} & \textbf{Watch} & \textbf{Combined} \\
				\midrule
				1 & 0.09 & 0.05 \\
				2 & 0.09 & 0.07 \\
				3 & 0.09 & 0.07 \\
				4 & 0.06 & 0.05 \\
				5 & 0.06 & 0.03 \\
				6 & 0.04 & 0.02 \\
				F & 0.05 & 0.04 \\
				\midrule
				agnostic & 0.09 & 0.05 \\
				\bottomrule
			\end{tabular}
			\label{tab:ResultsPositionsWatch}
		} \\
	\end{tabular}
	\caption{Average EERs for our \textit{terminal-specific} payment authentication models in optimum window $\{s=2.5, o=0\}$, using data from the smart ring only, the smartwatch only, and both combined. Our \textit{terminal-agnostic} results are included for comparison.} 
	\label{tab:ResultsPositions}
\end{table}

\subsection{Enrolment Parameters}

Behavioural biometric systems typically entail a burdensome enrolment phase, where the user must perform the measured characteristic repeatedly to create the initial template. To evaluate the extent to which we can expedite the enrolment phase, we compare the average EERs of our authentication models when the classifiers are trained on a smaller positive class (\textit{i.e.}, fewer user samples).

\begin{figure}[t!]
	\vspace{-1.2em}
	\centering
	\begin{tabular}{c}
		\subfloat[payment model in $\{s=2.5, o=0\}$]{
			\includegraphics[height=5.0cm]{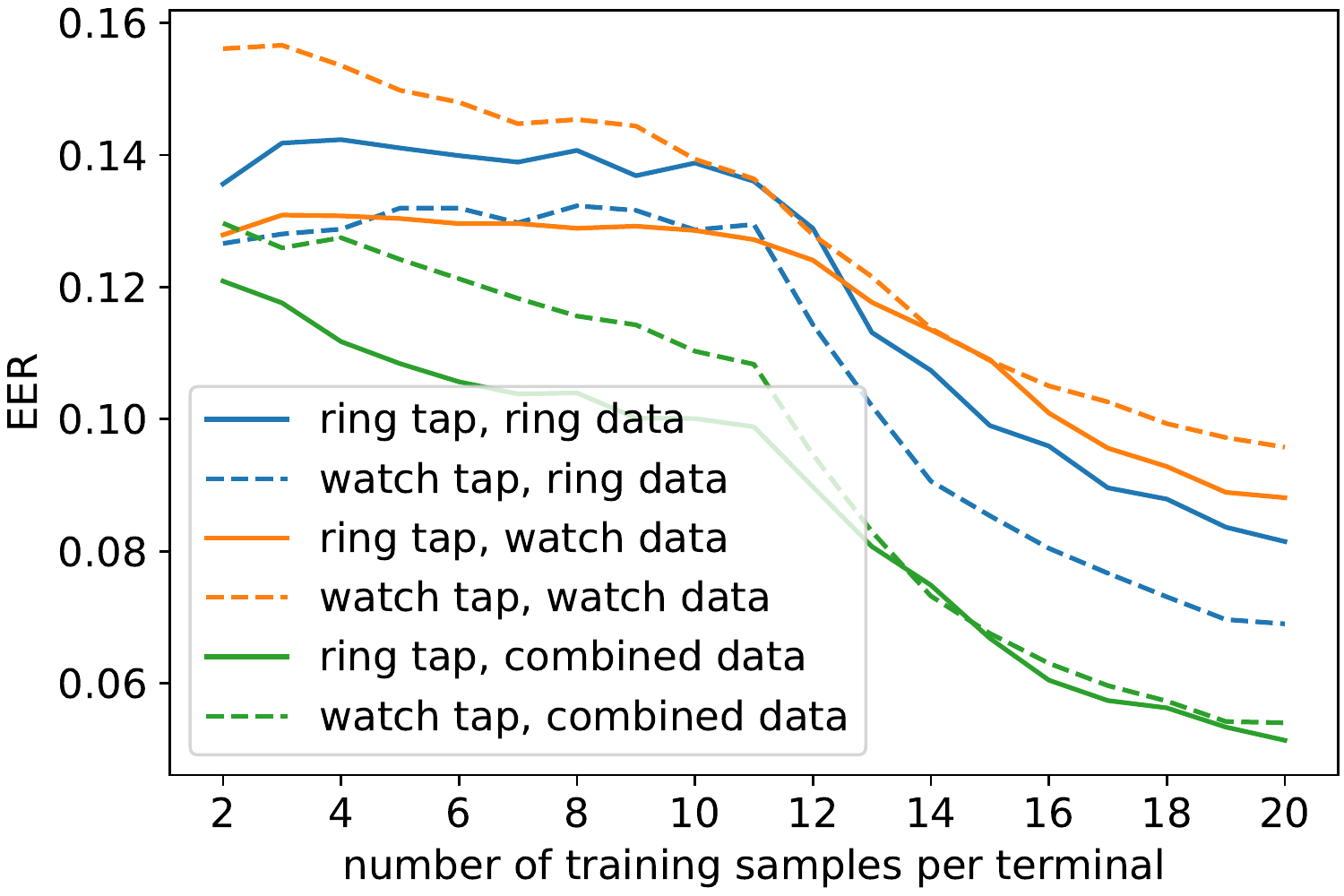}
			\label{fig:ResultsTrainingSizePayment}
		} \\
		\subfloat[access control model, 3-knock gesture]{
			\includegraphics[height=5.0cm]{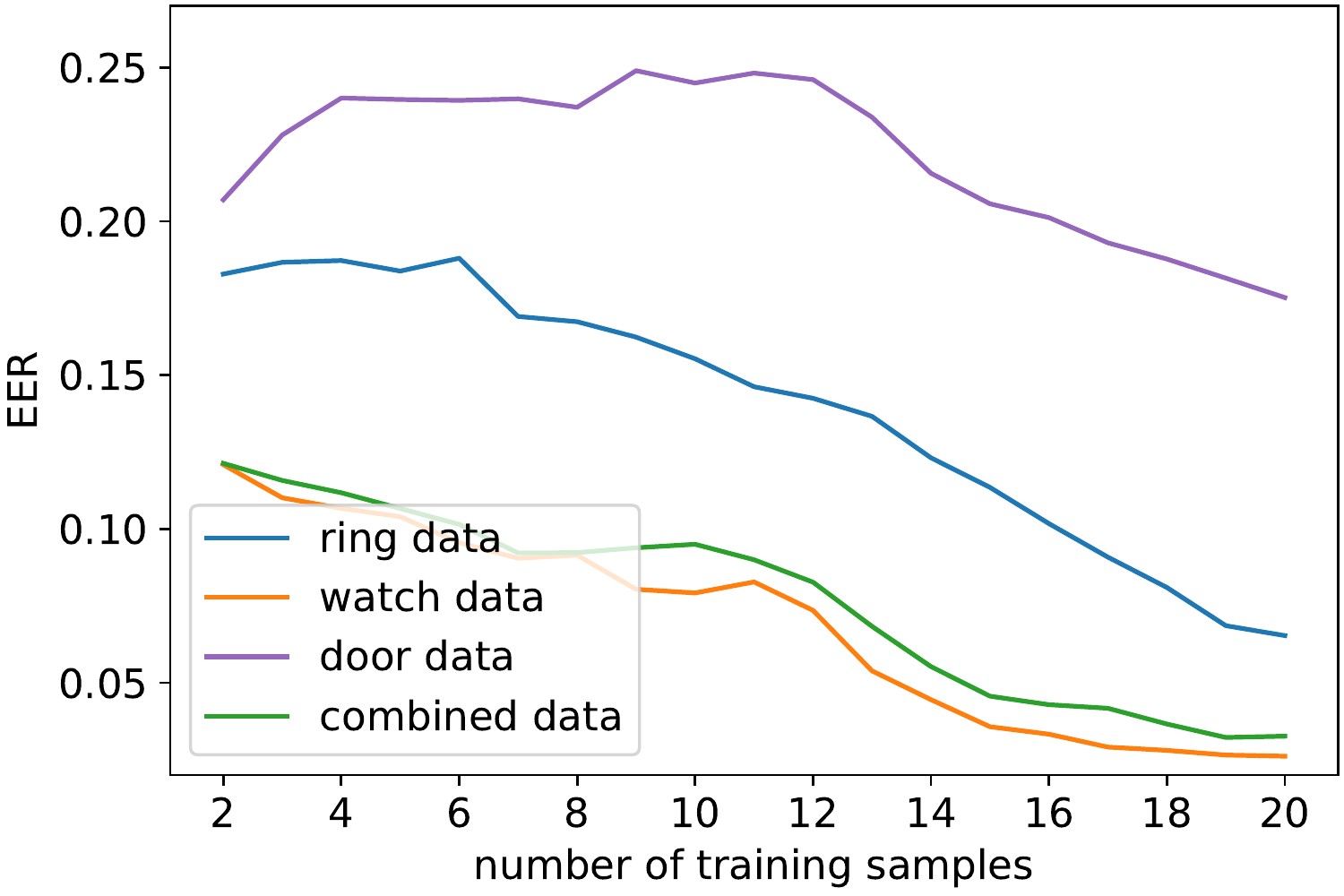}
			\label{fig:ResultsTrainingSizeDoor}
		} \\
	\end{tabular}
	\caption{Average EERs for our payment and access control models if trained on different numbers of enrolment samples. For the payment model, each classifier is trained on six terminals.}
	\label{fig:ResultsTrainingSize}
\end{figure}

Figure \ref{fig:ResultsTrainingSizePayment} shows that our payment models can authenticate the user with EERs as low as 0.12 when trained on just twelve of the user's tap gestures (spread evenly over six terminals), which can be performed in less than a minute. Figure \ref{fig:ResultsTrainingSizeDoor} shows that, when including watch data, our 3-knock access control model can authenticate the user with EERs as low as 0.12 when trained on just two 3-knock gestures (the access control models for the other knock gestures show a similar pattern), which can be performed in a few seconds. In both cases, we see that the EERs improve as more samples are included in the training set; this suggests that an update mechanism might benefit the models over time, relaxing upfront requirements and incorporating subsequent gestures as the system is used.

\section{Discussion}\label{sec:Discussion}

\textbf{Power Consumption.} Wearable devices are designed to facilitate always-on sensing (\textit{e.g.}, in health monitoring applications). To measure the impact of our data collection in practical terms, we wore two of each wearable device in an identical state, but only collecting data from one of each. For the smart rings, there was no noticable difference in power consumption over 6 hours. For the smartwatches, without any effort put into performance optimisation, our app caused the smartwatch running it to consume an additional 1.5\% of battery capacity per hour. While we did not implement the random forest classifier on the devices, we argue that its energy consumption would be negligible due to the limited number of inferences that would be required per day (only when the user needs to authenticate, such as to make a payment).

\textbf{Response Time.} We calculated the computation time for classifying a single watch tap gesture, averaged over 10,000, to be 7.11 ms for authentication on a desktop computer with an Intel Core i5-6500 processor. Using a benchmarking tool\footnote{https://www.notebookcheck.net}, we found that a Samsung Exynos W920 (a modern smartwatch processor) performs 26 times slower, so we would expect an authentication decision to be made in roughly 185 ms on a smartwatch. Robust benchmarking is not yet publicly available for smart ring CPUs.

\textbf{User Feedback.} All of the participants in our user study found ring tap gestures to be easier and more comfortable to perform than watch tap gestures, due to the manoeuvrability of the hand compared to that of the wrist; some terminal positions were more awkward than others, depending on the height of the user and the wrist upon which the smartwatch was worn. One female participant, who wore the devices on her right arm, commented that she considered the smartwatch used in this study to be a men's watch due to its bulk and that, while she would normally wear a women's (smaller) watch on her right (dominant) wrist, she would have to wear this one on her left wrist for daily use because she would find it obstructive otherwise.

\section{Related Work}\label{sec:RelatedWork}

\textbf{Tap Gestures.} The use of inertial sensor data to authenticate tap gestures in tap-and-pay systems was proposed by Shrestha \textit{et al.} \cite{Shrestha2016} for smartphone-based systems, achieving F-measure scores of 0.93, and by Sturgess \textit{et al.} \cite{Sturgess2022-1, Sturgess2022-2} for smartwatch-based systems, achieving F-measure scores of 0.87 and EERs of 0.08. The use of a smartwatch, due to the physiology of the arm, introduced the challenge that the sensor axes frequently change reference frames because the device changes orientation during the tap gesture. We found that the use of a smart ring sits between the two in terms of complexity: the smartphone requires no major change in orientation, the smart ring requires only a single change because the finger is easily manoeuvred towards the terminal, and the smartwatch orientation is changed frequently during the tap gesture. We found similar results in our smartwatch models and improved results with our smart ring and combined models.

\textbf{Smartwatches.} The use of inertial sensors on smartwatches have been used in a variety of authentication cases. Johnston \textit{et al.} \cite{Johnston2015} showed that wrist motion data can be used to both identify and authenticate a user while walking with 10-second windows of data. Nassi \textit{et al.} \cite{Nassi2016} showed that wrist motion data can be used to authenticate handwritten signatures and other authors \cite{Ciuffo2017, Griswold-Steiner2017, Griswold-Steiner2019, Wijewickrama2021} applied a similar approach to freestyle handwriting with 5- to 60-second windows of data. We found optimum results with 2.5 seconds of data for tap gestures and 2.8 seconds for knock gestures.

A number of works have used inertial sensor data from a smartwatch to support the authentication of a user on another device. Mare \textit{et al.} \cite{Mare2014} showed that wrist motion data can be used to infer a sequence of interactions from a user and correlated against inputs on his workstation, such that he can be de-authenticated if the correlation stops (however, the system was found to have vulnerabilities due to design flaws \cite{Huhta2016}). Acar \textit{et al.} \cite{Acar2020} showed that wrist motion data can be correlated with keystrokes to continuously authenticate the user of the workstation. Other works \cite{Lee2016, Mare2019} have correlated wrist motion data with smartphone interactions to authenticate the user of the smartphone.

\textbf{Smart Rings.} Few authors have considered the use of smart rings in authentication use-cases. Sen \textit{et al.} \cite{Sen2020} proposed the use of a smart ring that is capable of producing a vibration to bootstrap a communication channel with another device held in the same hand that has an accelerometer to detect the vibration. Liang \textit{et al.} \cite{Liang2017} showed that inertial sensor data from a smart ring can be correlated with mouse movements to continuously authenticate the user of a workstation. To the best of our knowledge, we are the first to propose the use of inertial sensors on a smart ring to authenticate a user via implicit or explicit gesture biometrics.

\section{Limitations and Future Work}\label{sec:LimitationsandFutureWork}

\textbf{3-knock Impersonation.} Table \ref{tab:ResultsDoor} shows that we achieved our best overall results from the 3-knock gesture. In our preliminary experiments, this was not the case, so we chose not to include it in the impersonation exercise of our user study. This is regrettable, as the observation attack may have yielded more interesting results for 3-knock gestures than for 5-knock gestures.

\textbf{Sampling Rate.} By comparing Figure \ref{fig:ResultsDown} with Figure \ref{fig:ResultsUndown} in the Appendix, we see that downsampling our smart ring data from 100 Hz to 50 Hz imposed only a slight cost in performance (at most, a difference of 0.01 in average EERs). Nonetheless, future work should endeavour to use a smartwatch and Raspberry Pi IMU with higher sampling rates, to remove any reason for downsampling, and may see improved scores across the board.

\section{Conclusion}\label{sec:Conclusion}

In this paper, we showed that inertial sensor data from a smart ring can be used to authenticate the wearer. In a mobile payment context, we showed that a smart ring user can be implicitly authenticated with a single tap gesture with an EER of 0.04. We also showed that inertial sensor data from a smart ring can be used to authenticate the user when making a smartwatch payment, and \textit{vice versa}, opening the possibility for either device to be used as an implicit second factor for the other. In an access control context, we showed that a smart ring user can be (explicitly) authenticated with a single knock gesture with an EER of 0.06 (or 0.02 with a smartwatch). We demonstrated that our authentication models provide resistance against an active impersonation attacker who observed the victim's gestures and we showed that successful attacks were more likely the result of luck than of a skilled attacker.

\section*{Acknowledgement}

This work was supported financially by Mastercard; the Engineering and Physical Sciences Research Council [grant number EP/P00881X/1]; and the PETRAS National Centre of Excellence for IoT Systems Cybersecurity [grant number EP/S035362/1]. The authors would like to thank these organisations for their support, Genki Instruments for their collaboration and technical support, and the anonymous reviewers for their feedback.

\setcounter{figure}{0}
\renewcommand{\thefigure}{A.\arabic{figure}}
\section*{Appendix}

\begin{figure}[h!]
	\vspace{-1em}
	\centering
	\begin{tabular}{ccc}
		\hspace{3.4em}
		\subfloat[ring tap; ring data]{
			\includegraphics[height=2.6cm]{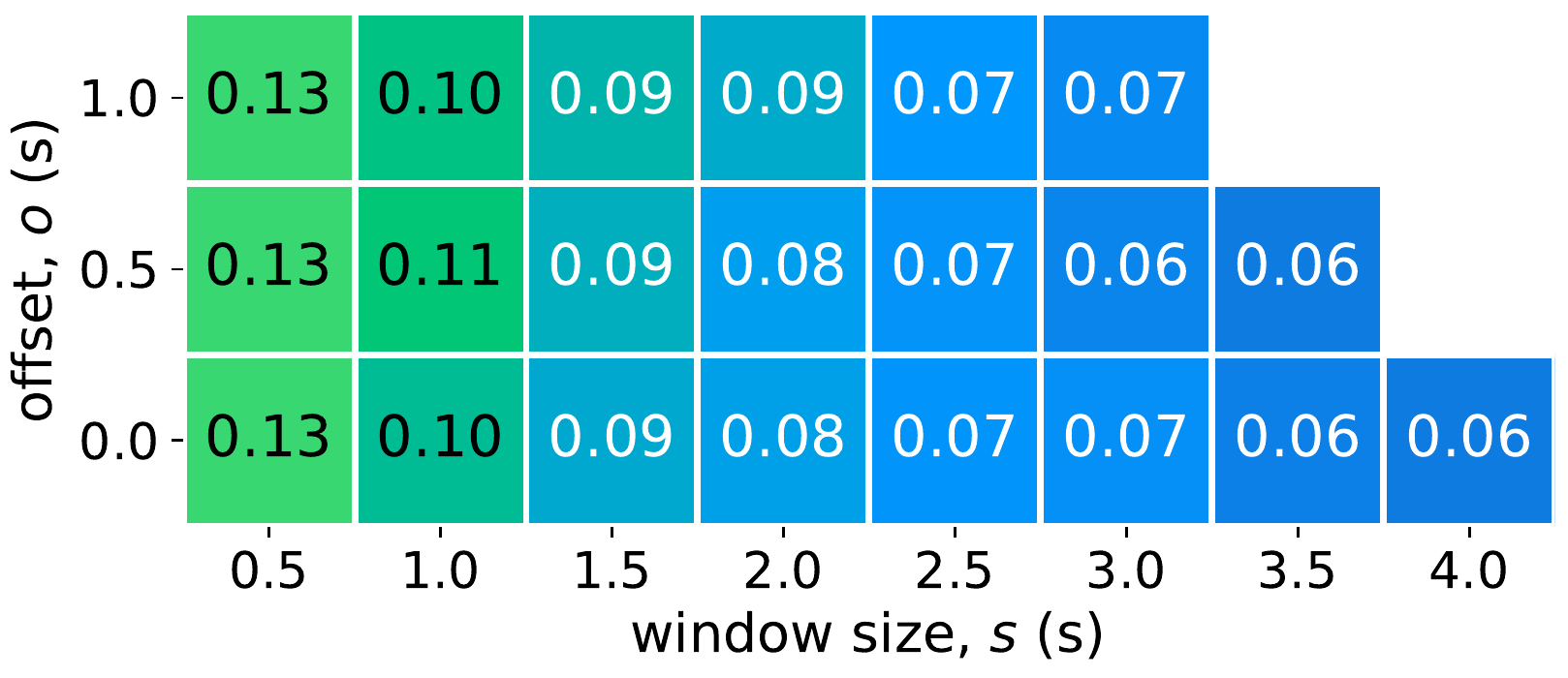}
			\label{fig:ResultsUndownRingRing}
		} & 
		\subfloat[watch tap; ring data]{
			\includegraphics[height=2.6cm]{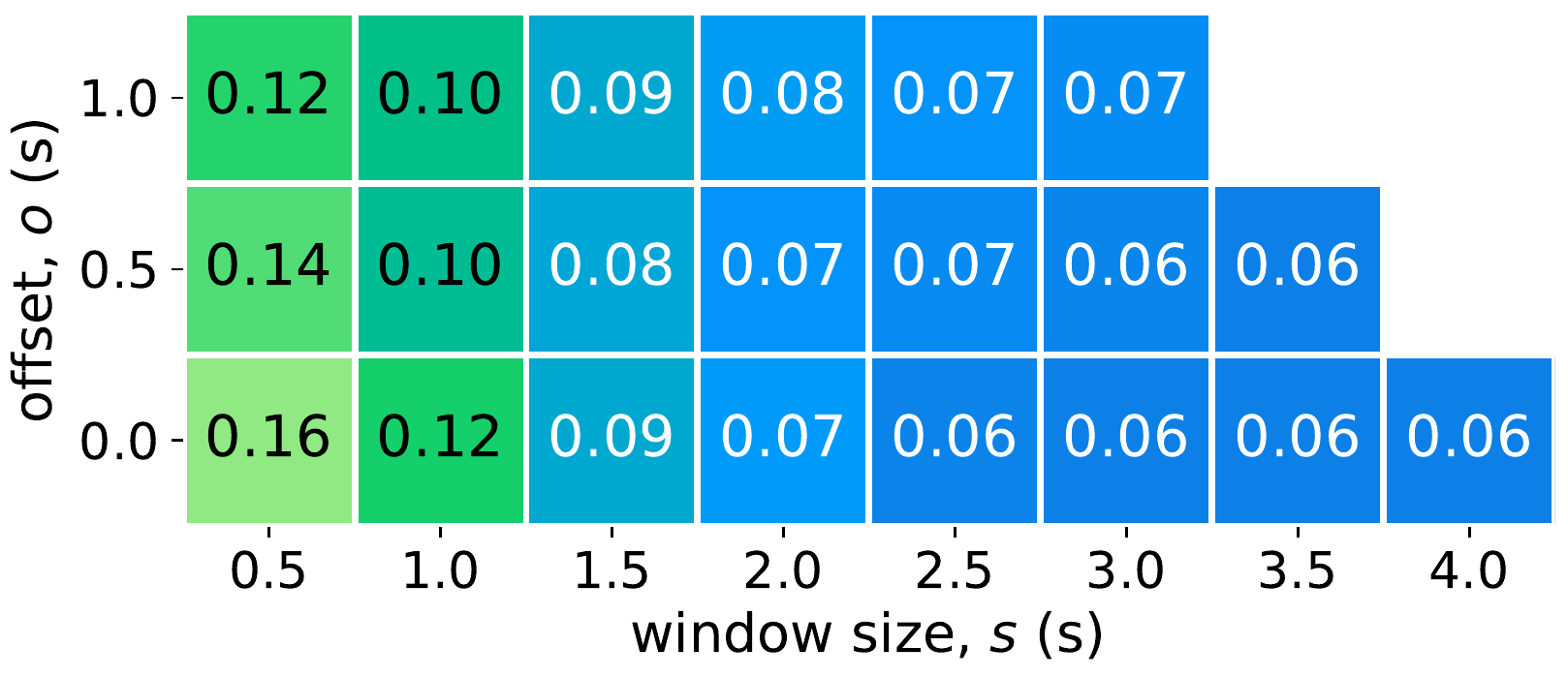}
			\label{fig:ResultsUndownWatchRing}
		} & 
		\multirow{-7}[2]{*}{\subfloat{\includegraphics[height=6.0cm]{figures/legends_eer}}} \\
		\setcounter{subfigure}{2}%
		\hspace{3.4em}
		\subfloat[ring tap; combined data]{
			\includegraphics[height=2.6cm]{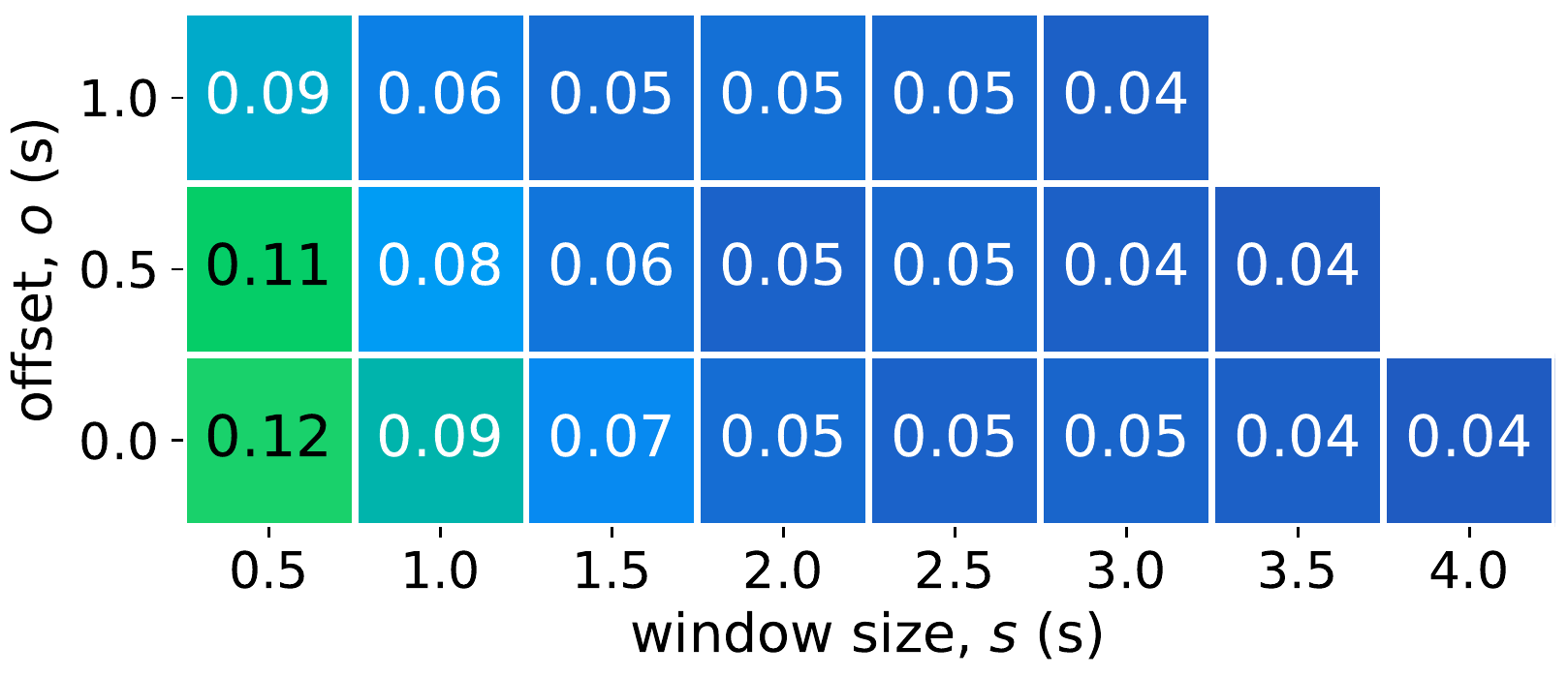}
			\label{fig:ResultsUndownRingCombined}
		} & 
		\subfloat[watch tap; combined data]{
			\includegraphics[height=2.6cm]{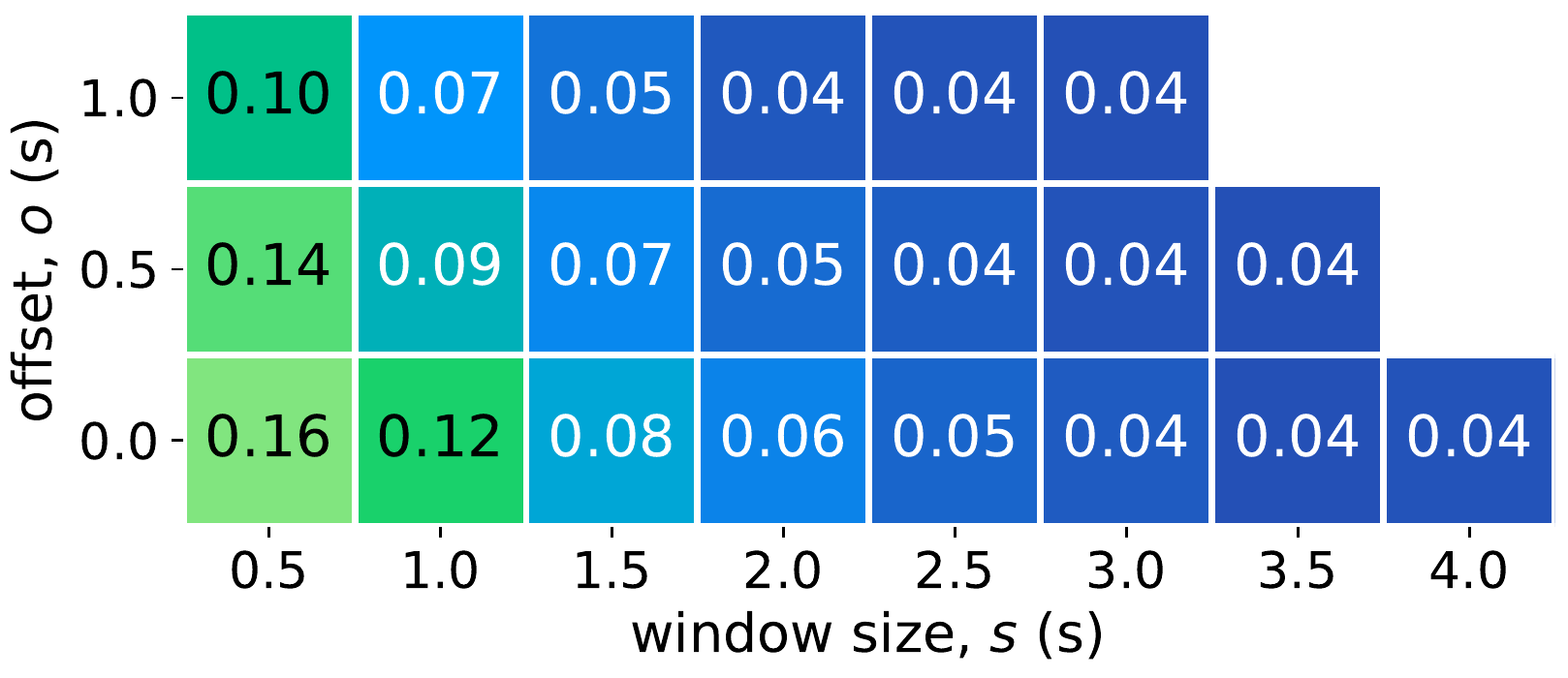}
			\label{fig:ResultsUndownWatchCombined}
		} \\
	\end{tabular}
	\caption{Average EERs for our payment authentication models by window size and offset, for tap gestures made with the smart ring (left) and smartwatch (right), using data from the smart ring only (top) and the smart ring and smartwatch combined (bottom), in all cases based on undownsampled data from the smart ring. (\textit{cf.} Figure \ref{fig:ResultsDown} to compare with the downsampled smart ring data.)}
	\label{fig:ResultsUndown}
\end{figure}

\begin{figure}[h!]
	\vspace{-1em}
	\centering
	\begin{tabular}{ccc}
		\hspace{3.4em}
		\subfloat[ring tap; watch data]{
			\includegraphics[height=2.6cm]{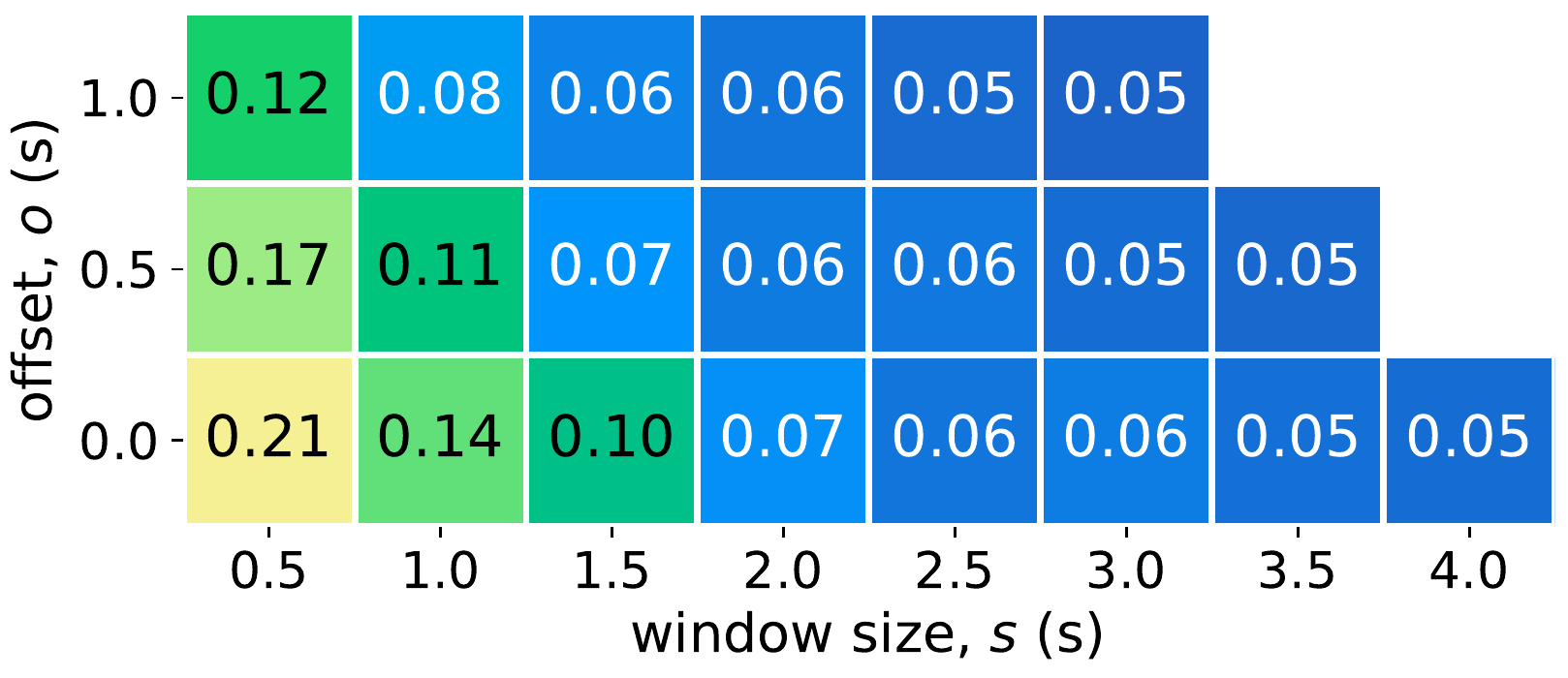}
			\label{fig:ResultsDownAccGyrRingWatch}
		} & 
		\subfloat[watch tap; watch data]{
			\includegraphics[height=2.6cm]{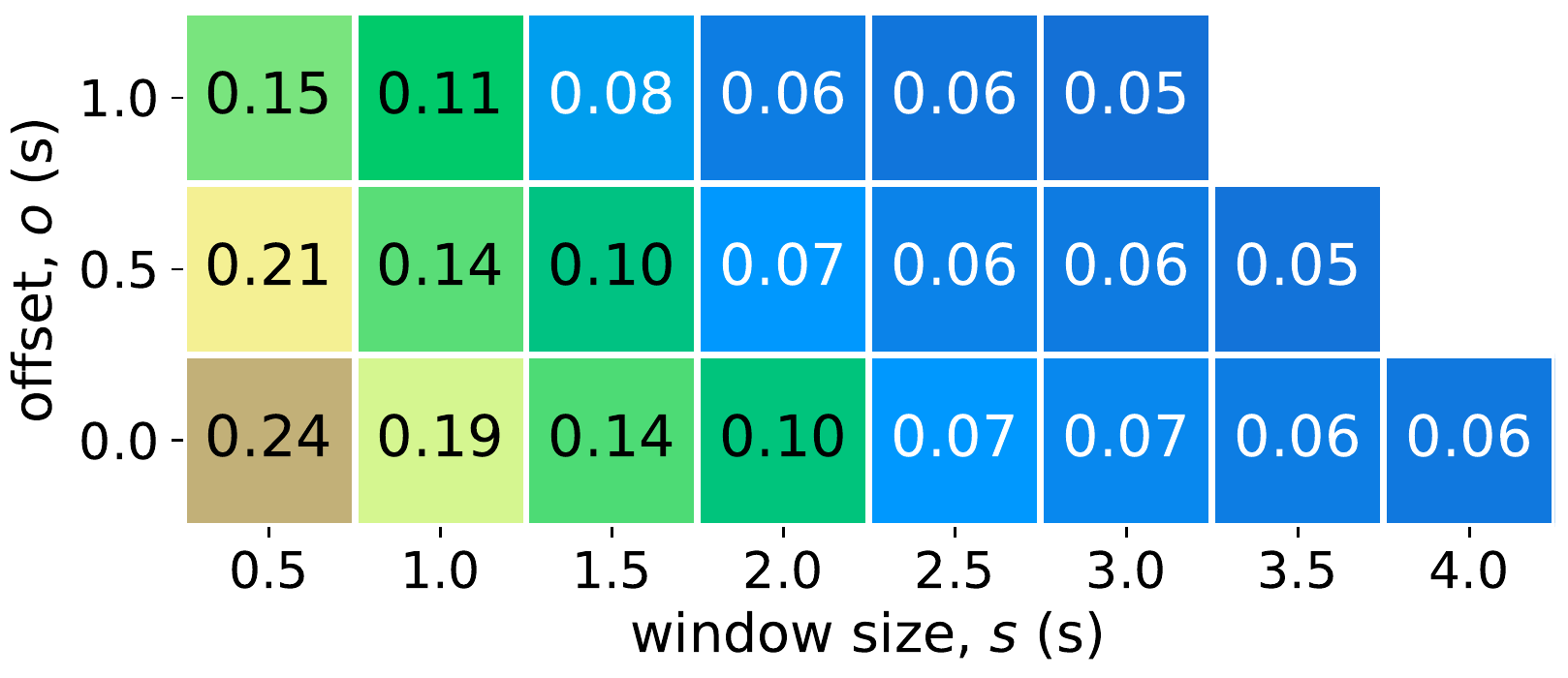}
			\label{fig:ResultsDownAccGyrWatchWatch}
		} & 
		\subfloat{\includegraphics[height=6.0cm]{figures/legends_eer}} \\
	\end{tabular}
	\caption{Average EERs for our payment authentication models by window size and offset, for tap gestures made with the smart ring (left) and smartwatch (right), using only accelerometer and gyroscope data from the smartwatch. (\textit{cf.} Figure \ref{fig:ResultsDown} to compare with the all-sensor data.)}
	\label{fig:ResultsDownAccGyr}
\end{figure}


\begin{thebibliography}{00}

\bibitem{Acar2020}
A. Acar, H. Aksu, A. S. Uluagac, and K. Akkaya.
``A Usable and Robust Continuous Authentication Framework using Wearables",
\textit{IEEE Transactions on Mobile Computing (TMC)}, 2020.

\bibitem{Ciuffo2017}
F. Ciuffo and G. M. Weiss.
``Smartwatch-based Transcription Biometrics",
\textit{IEEE Annual Ubiquitous Computing, Electronics \& Mobile Communication Conference}, 2017.

\bibitem{Griswold-Steiner2017}
I. Griswold-Steiner, R. Matovu, and A. Serwadda.
``Handwriting Watcher: A Mechanism for Smartwatch-driven Handwriting Authentication",
\textit{IEEE International Joint Conference on Biometrics (IJCB)}, 2017.

\bibitem{Griswold-Steiner2019}
I. Griswold-Steiner, R. Matovu, and A. Serwadda.
``Wearables-driven Freeform Handwriting Authentication",
\textit{IEEE Transactions on Biometrics, Behavior, and Identity Science}, Vol. 1, 2019.

\bibitem{Huhta2016}
O. Huhta, P. Shrestha, S. Udar, M. Juuti, N. Saxena, and N. Asokan.
``Pitfalls in Designing Zero-effort Deauthentication: Opportunistic Human Observation Attacks",
\textit{Network and Distributed System Security Symposium (NDSS)}, 2016.

\bibitem{Johnston2015}
A. H. Johnston and G. M. Weiss.
``Smartwatch-based Biometric Gait Recognition",
\textit{IEEE International Conference on Biometrics Theory, Applications, and Systems (BTAS)}, 2015.

\bibitem{Lee2016}
W. H. Lee and R. B. Lee.
``Implicit Sensor-based Authentication of Smartphone Users with Smartwatch",
\textit{ACM Hardware and Architectural Support for Security and Privacy (HASP)}, 2016.

\bibitem{Liang2017}
X. Liang and D. Kotz.
``AuthoRing: Wearable User-presence Authentication",
\textit{ACM Workshop on Wearable Systems and Applications (WearSys)}, 2017.

\bibitem{Mare2014}
S. Mare, A. M. Markham, C. Cornelius, R. Peterson, and D. Kotz.
``ZEBRA: Zero-Effort Bilateral Recurring Authentication",
\textit{IEEE Symposium on Security and Privacy (S\&P)}, 2014.

\bibitem{Mare2019}
S. Mare, R. Rawassizadeh, R. Peterson, and D. Kotz.
``Continuous Smartphone Authentication using Wristbands",
\textit{Workshop on Usable Security and Privacy (USEC)}, 2019.

\bibitem{Nassi2016}
B. Nassi, A. Levy, Y. Elovici, and E. Shmueli.
``Handwritten Signature Verification using Hand-worn Devices",
\textit{ACM Interactive, Mobile, Wearable and Ubiquitous Technologies (IMWUT)}, Vol. 2, 2016.

\bibitem{Sen2020}
S. Sen and D. Kotz.
``VibeRing: Using Vibrations from a Smart Ring as an Out-of-band Channel for Sharing Secret Keys",
\textit{Journal of Pervasive and Mobile Computing}, Vol. 78, 2020.

\bibitem{Shrestha2016}
B. Shrestha, M. Mohamed, S. Tamrakar, and N. Saxena.
``Theft-Resilient Mobile Wallets: Transrgessparently Authenticating NFC Users with Tapping Gesture Biometrics",
\textit{Annual Conference on Computer Security Applications (ACSAC)}, 2016.

\bibitem{Sturgess2022-1}
J. Sturgess, S. Eberz, I. Sluganovic, and I. Martinovic.
``Inferring User Height and Improving Impersonation Attacks in Mobile Payments using a Smartwatch",
\textit{IEEE International Conference on Pervasive Computing and Communication Workshops (PerCom Workshops)}, 2022.

\bibitem{Sturgess2022-2}
J. Sturgess, S. Eberz, I. Sluganovic, and I. Martinovic.
``WatchAuth: User Authentication and Intent Recognition in Mobile Payments using a Smartwatch",
\textit{IEEE European Symposium on Security and Privacy (EuroS\&P)}, 2022.

\bibitem{Wijewickrama2021}
R. Wijewickrama, A. Maiti, and M. Jadliwala.
``Write to Know: On the Feasibility of Wrist Motion Based User-Authentication from Handwriting",
\textit{ACM Conference on Security and Privacy in Wireless and Mobile Networks (WiSec)}, 2021.

\end{thebibliography}
\end{document}